\documentclass[twocolumn]{aastex6}

\usepackage{amsmath}

\usepackage{units}

\usepackage{hyperref}

\begin{document}
\label{firstpage}

\shorttitle{3D simulations of wind mass transfer in binaries}
\title{Three-dimensional hydrodynamical simulations of mass transfer in binary systems by a free wind}

\shortauthors{Liu Z.-W. et al.}
\author{Zheng-Wei Liu\altaffilmark{1}, Richard J. Stancliffe\altaffilmark{1}, Carlo Abate\altaffilmark{1} and Elvijs Matrozis\altaffilmark{1}}

\altaffiltext{1}{Argelander-Institut f\"ur Astronomie, Auf dem H\"ugel 71, D-53121, Bonn, Germany}

\email{Email: zwliu@ynao.ac.cn}

\begin{abstract}

A large fraction of stars in binary systems are expected to undergo mass and angular momentum exchange at some point in their evolution, which can drastically alter the chemical and dynamical properties and fates of the systems. Interaction by stellar wind is an important process in wide binaries. However, the details of wind mass transfer are still not well understood. We perform three-dimensional hydrodynamical simulations of wind mass transfer in binary systems to explore mass accretion efficiencies and geometries of mass outflows, for a range of mass ratios from $0.05$ to $1.0$. In particular, we focus on the case of a free wind, in which some physical mechanism accelerates the expelled wind material balancing the gravity of the mass-losing star with the wind velocity comparable to the orbital velocity of the system. We find that the mass accretion efficiency and accreted specific angular momentum increase with the mass ratio of the system. For an adiabatic wind, we obtain that the accretion efficiency onto the secondary star varies from about 0.1\% to 8\% for mass ratios between $0.05$ and $1.0$.

\end{abstract}


\section{INTRODUCTION}
 \label{sec:introduction}

A large fraction of stars are in binary or multiple systems \citep{Duquennoy1991, Raghavan2010}. Stars in interacting binary systems undergo mass and angular momentum exchange through the transfer of material between the two stars, and the system as a whole may lose both mass and angular momentum when material is ejected from the system. Binary interactions can strongly alter the properties of the accreting companion star, for example its spin and chemical composition of the outer layers, and the orbital parameters of the binary system. Such interactions are therefore expected to play an essential role in a wide range of astronomical phenomena associated with binary stars, e.g., X-ray binaries, carbon-enhanced metal-poor (CEMP) stars, Type Ia supernovae (SNe Ia), microquasars, and asymmetric planetary nebulae (PNe).

Mass transfer in a close binary system usually occurs through Roche lobe overflow (RLOF), in which the primary star starts to transfer material to its companion star once it fills its Roche lobe \citep{Kopal1959, Paczynski1971, Eggleton1983}. However, many observed interacting binaries are too wide to undergo RLOF. In these systems mass transfer is believed to occur via the stellar wind. The consequences of wind accretion are quite different from those of RLOF. For instance, the accretion efficiency of the mass-transfer process is expected to be lower than in the case of stable RLOF. The bulk of the mass can in fact be lost from the system as wind accretion occurs, while in the RLOF case nearly all of the transferred material can be accreted onto the companion star \citep{Hoyle1939, Pringle1985}.

Wind mass transfer in a binary system,  consisting of a mass-losing asymptotic giant branch (AGB) star and an accreting main-sequence companion, is widely believed to be the formation scenario of chemically-peculiar stars such as barium stars \citep{Smith1996} and CEMP stars \citep{Wallerstein1998, Stancliffe2007, Stacliffe2008}. Also, symbiotic systems, i.e. systems in which a white dwarf accretes material from a mass-losing red giant (or AGB) companion star, have been suggested to eventually produce SNe Ia \citep{Hachisu1999, Claeys2014}. Therefore, addressing the details of wind accretion in binary systems is of fundamental importance for a better understanding of these astrophysical phenomena.

Wind accretion was first studied by \citet{Hoyle1939, Bondi1944} and \citet{Bondi1952} who developed the so-called Bondi-Hoyle-Lyttleton (BHL) model of this process (see \citealt{Edgar2004} for a recent review). Their prescription is typically used for wind mass transfer in binary population synthesis calculations (e.g., \citealt{Izzard2009, Pols2012, Claeys2014}). Many numerical simulations have been performed to study wind accretion and test the validity of the predictions of the canonical BHL model of wind mass transfer (e.g., \citealt{Hunt1971, Anzer1987, Livio1986, Bowen1991, Matsuda1992, Sawada1992, Ruffert1994a, Bate1995}). As discussed by \citet{de-Val-Borro2009}, for high-velocity winds, the mass-accretion rate in binary systems can be reasonably described by BHL accretion, which assumes that the wind speed is much faster than the orbital speed of the accretor. However, AGB stars are expected to have slow winds which are in many cases of the same order of magnitude as their orbital velocities, i.e., typically $5$--$30\,\mathrm{km\,s^{-1}}$ \citep{Vassiliadis1993, Knapp1998}. This makes the description of wind accretion in symbiotic systems much more complex than in the fast-wind approximation. Consequently, the canonical BHL accretion cannot adequately explain the observations of the nearest and best-studied symbiotic binary, Mira, characterized by slow wind speeds \citep{Karovska1997, Karovska2005}. Furthermore, several hydrodynamical simulations in two and three dimensions have found that the mass-accretion rate is very different from that estimated by the BHL when the wind speed is comparable to (or slower than) the orbital velocity (e.g., \citealt{Bowen1988, Theuns1993, Theuns1996, Ruffert1994a, Bate1995, Mastrodemos1998, Nagae2004, Jahanara2005, Mohamed2007, Mohamed2012, de-Val-Borro2009}). For instance, \citet{Theuns1996} found that the mass-accretion rate in the model with a ratio of specific heats of $\gamma$=$1.5$ is about ten times smaller than theoretical estimates based on the BHL prescription. On the other hand, \citet{Mohamed2012} suggest a new mode of mass transfer somewhere in between RLOF and wind mass transfer, called the wind Roche-lobe overflow (WRLOF), which could lead to much more efficient mass accretion onto the secondary compared to canonical BHL accretion.

Recently, \citet{deValBorro2017} found that mass accretion rates during outburst-related accretion in symbiotic systems are 20--50\% larger than those in the standard BHL approximation. Also, simulations for AGB binary systems by \citet{Chen2017} suggested that the mass-transfer in wide binaries is characterised by the BHL accretion, while their shortest-orbit binary system experiences WRLOF mass transfer and has more efficient mass accretion, consistent with the results of \citet{Mohamed2010, Mohamed2011, Mohamed2012}.

The key questions that still need to be addressed in the context of wind mass transfer in wide binary systems are:
\begin{itemize}\itemsep-2.5pt
\item[-]{How efficiently are mass and angular momentum transferred by stellar winds?}
\item[-]{How do mass and angular momentum accreted by the secondary and lost from the system vary with the mass ratio and orbital separation of the binary?}
\item[-]{How does the eccentricity affect accretion properties?}
\item[-]{How does the presence of an accreting companion star alter the structure of the stellar wind?}
\end{itemize}

In this work, we perform three-dimensional (3D) Smoothed Particle Hydrodynamics (SPH) modelling to investigate the wind mass transfer in the case of wind velocity comparable to the orbital velocity of the system. The main purpose of this work is to quantify the mass and angular momentum that are accreted by the secondary stars and in particular to investigate how the results vary as a function of the mass ratio  of the binary systems ($q$=$M_{2}/M_{1}\leqslant 1$). In the present work we extend the study of \citet{Theuns1993} to cover a wider range of the mass ratios of the systems from 0.05 to 1.0. We defer to forthcoming papers the study of other physical parameters, such as the ratio of wind velocity over orbital velocity ($\upsilon_{\mathrm{wind}}$/$\upsilon_{\mathrm{orb}}$), the orbital separation ($A_{\mathrm{orb}}$), and the orbital eccentricity ($e$), which are expected to have a major influence on accretion properties of slow-wind accretion in binary systems \citep{Nagae2004, Jahanara2005, de-Val-Borro2009, deValBorro2017}.

\begin{figure*}
  \begin{center}
    {\includegraphics[width=0.48\textwidth, angle=360]{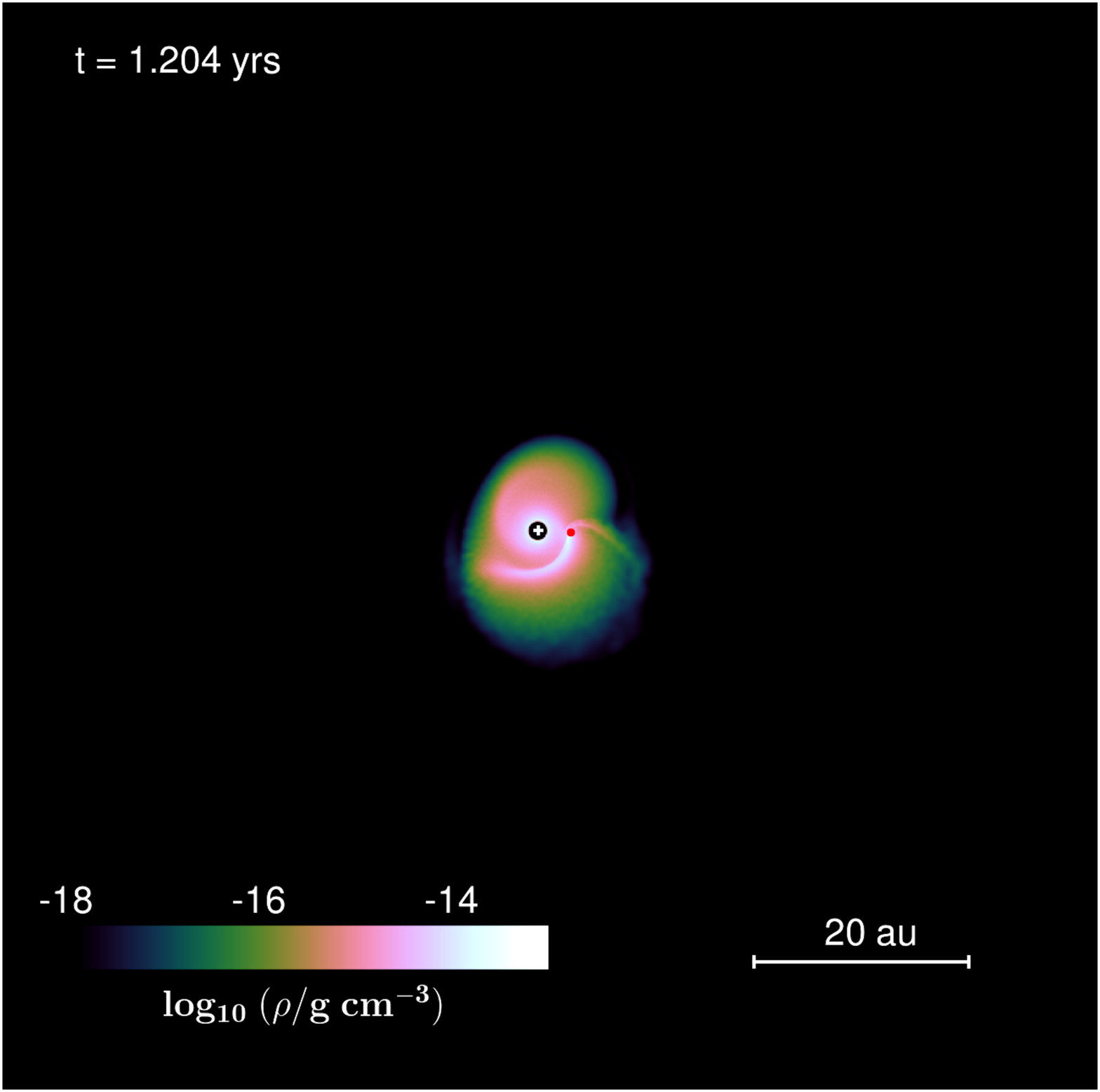}}
    {\includegraphics[width=0.48\textwidth, angle=360]{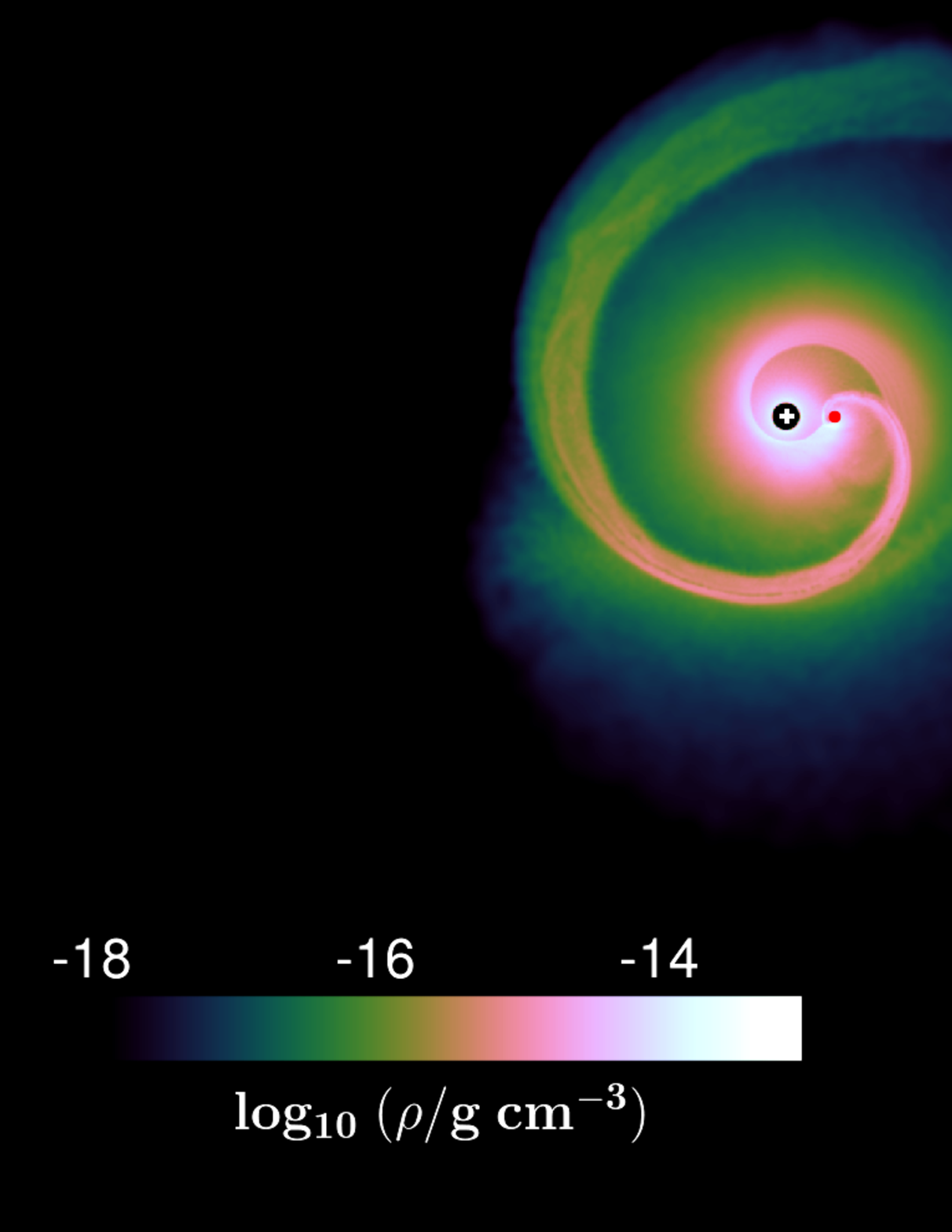}}
    {\includegraphics[width=0.48\textwidth, angle=360]{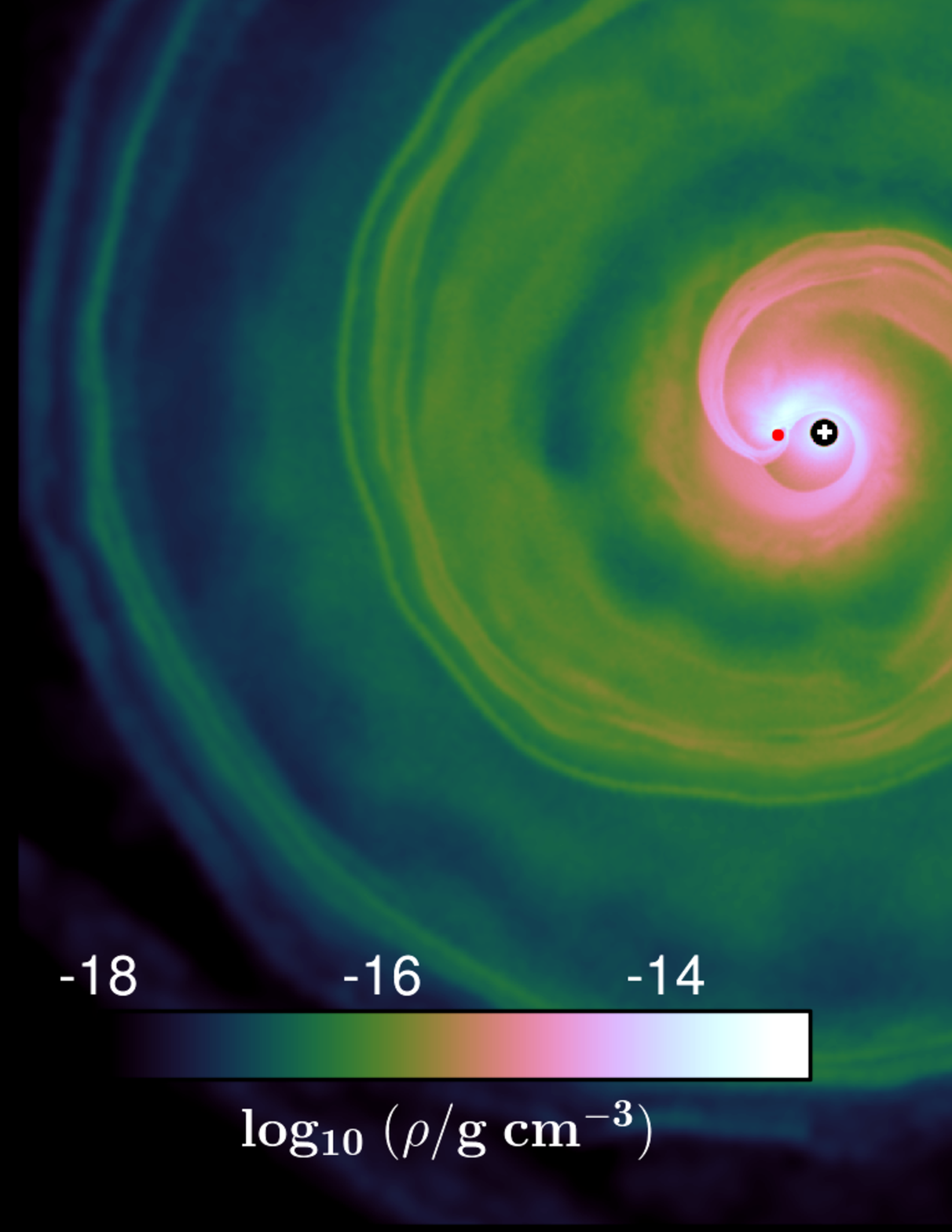}}
    {\includegraphics[width=0.48\textwidth, angle=360]{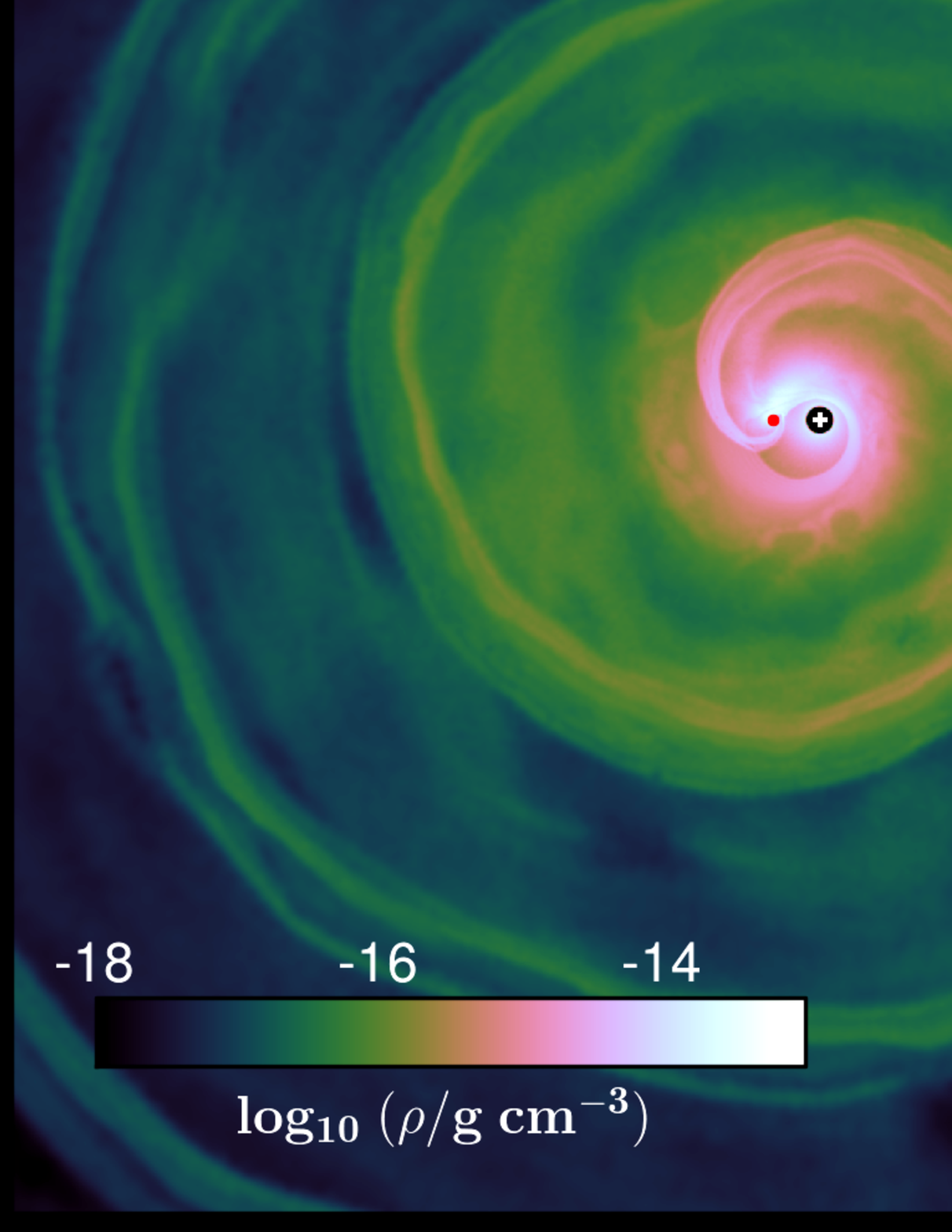}}
  \caption{Cross-section slice of gas density in the $x-y$ plane  ($z=0$) in the simulation of a binary system consisting of a $3.0\,M_{\sun}$ mass-losing star (white cross) and a $1.5\,M_{\sun}$ accreting companion (red dot). The binary system has an orbital separation of $3\,\rm{au}$ (which corresponds to an orbital period of about $2.45\,\mathrm{yrs}$). The mass-losing star has a radius of $200\,R_{\sun}$ and it is losing mass via stellar winds at a rate of $10^{-6}\,M_{\sun}\,\mathrm{yr^{-1}}$ with a wind velocity of $15\,\mathrm{km\,s^{-1}}$. All parameters of the binary system are set up to be consistent with \citet{Theuns1996}. The color scale shows the logarithm of the mass density in $\mathrm{g\,cm^{3}}$.}
\label{Fig:xy1}
  \end{center}
\end{figure*}

\section{Numerical Methods and Models}
\label{sec:method}

In this work, 3D simulations of wind mass transfer in binary systems are performed with the SPH method \citep{Lucy1977, Gingold1977, Benz1990}. SPH is a mesh-free Lagrangian method (where the coordinates move with the fluid), treating hydrodynamics in terms of a set of sampling particles. The gas properties on each particle in SPH are calculated by averaging over its nearest neighbours. As a major advantage, the particle representation of SPH leads to an excellent conservation of energy, linear momentum, angular momentum and mass in simulations \citep{Springel2005}. Therefore, the SPH method is ideally suited to the simulation of dynamic phenomena which have arbitrary geometries and large deformations such as is the case in our simulation, although it suppresses hydrodynamical instabilities. We refer the reader to \citet{Rosswog2009} and \citet{Springel2010} for a detailed introduction of SPH  in the astrophysical context. In fact, many of the studies mentioned above have successfully applied the SPH method to model the stellar winds of AGB stars and wind mass transfer in binary systems (e.g., \citealt{Bowen1988, Theuns1993, Theuns1996, Bate1995, Hoefner2003, Woitke2006, Mohamed2012, Toupin2015}).

In our study we use the MPI-parallelised Lagrangian SPH code {\sc GADGET} \citep{Springel2005}. The {\sc GADGET} code uses an explicit communication model that is implemented with the standardized MPI communication interface. Originally, the {\sc GADGET} code was intended for cosmological simulations, but it has been modified to make it applicable to stellar astrophysics problems \citep{Pakmor2012, Liu12, Liu2013}. In this work we use the third version of the {\sc GADGET} code, {\sc GADGET-3}\footnote{This version is not publicly available. But the second version of {\sc GADGET}, {\sc GADGET2} \citep{Springel2005}, is freely available on \url{http://wwwmpa.mpa-garching.mpg.de/gadget/}.}, which has been modified to include binary motion, stellar winds and accretion for our wind accretion simulations.  Details on how we implemented the physics of wind mass loss and wind accretion in our code and on our most important assumptions are described in the next sections.

\subsection{Initial Setup}
\label{sec:setup}

We simulate a binary system consisting of an AGB star and a low mass accreting companion which is either a main sequence star or a white dwarf star in a circular orbit (i.e., orbital eccentricity $e$=$0$). In such binary system, an AGB star is losing mass by a stellar wind at a rate of about $10^{-6}\,M_{\sun}\,\rm{yr^{-1}}$. The binary systems of our simulations are set to orbit in the $x-y$ plane of a Cartesian coordinate system. To limit the computational cost of each simulation to a reasonable run time, the core of the AGB star and the detailed structure of the accreting star are not modelled. Instead, they are replaced by two non-gaseous, gravitation-only, massive particles. Here, we focus on modelling the wind structure in detail, we do not expect that detailed structures of two stars will significantly affect the results in our simulations. As a consequence, our simulation do not distinguish between main-sequence and white-dwarf companion stars. Treating the core of the AGB and the accreting star as two pure gravity sources enables us to only include their gravitational attractions but without any hydrodynamic interactions with other particles, i.e., these two non-gaseous particles have no effect on the dynamics of the simulation. The numerical resolution used in the majority of our simulations corresponds to a particle mass of $M_{\rm{SPH}}\approx3.6\times10^{-12}\,M_{\sun}$ unless stated otherwise.

\subsection{Wind Model}
\label{sec:wind}

The AGB wind is simulated by injecting  a number of SPH particles at a distance of $d$=$200\,R_{\sun}$ from the centre of the star to provide a constant mass-loss rate, $\dot{M}_{\mathrm{wind}}$, at each time interval, $\Delta\,t_{\rm{wind}}$. All wind particles are set up to have the same mass of $M_{\rm{SPH}}\approx3.6\times10^{-12}\,M_{\sun}$. As mentioned above, a constant mass-loss rate of $\dot M_{\rm{wind}}$=$10^{-6}\,M_{\sun}\,\rm{yr^{-1}}$ is assumed in the simulations. Therefore, the total number of SPH particles, $N_{\rm{insert}}$, that are injected into the simulation within a time interval ($\Delta\,t_{\rm{wind}}$) is $N_{\rm{insert}}$=$\Delta\,t_{\rm{wind}}\times\dot M_{\rm{wind}}/M_{\rm{SPH}}$. We inject wind particles isotropically in an annulus  around the mass-losing star. This setup enables us to model the wind structure in detail.

It is very important to ensure that artificial perturbations and discontinuities are minimised when new SPH particles are inserted. This requires distributing the inserted SPH particles smoothly and randomly within the shell. To create this desired condition, we adopt the method described by \citet{Springel2005} in his Section~5.4 to generate a ``glass-like'' particle distribution that naturally has zero pressure, as follows. First, collisionless particles are randomly distributed in a periodic box. This particle distribution is evolved according to the equations of motion  which include a fictitious repulsive gravitational force that relaxes the particles into a ``glass-like state''. Once a glass-like particle distribution in a periodic box has been created, a spherical shell is cut out from this distribution. Then, while keeping the azimuthal and polar positions of all collisionless particles of this ``cut-out'' shell constant, we rescale their radii to an appropriate value to reach the desired density distribution of a shell ($\rho \propto 1/r^{\,2}$, see \citealt{Ivezic1997}). Finally, all fluid characteristics (position, density, temperature and velocity) are assigned to these collisionless particles, making them become fluid SPH particles. These spherical shells are then given a random orientation to avoid producing artifacts in the simulation.

\begin{figure}
  \begin{center}
    {\includegraphics[width=0.49\textwidth, angle=360]{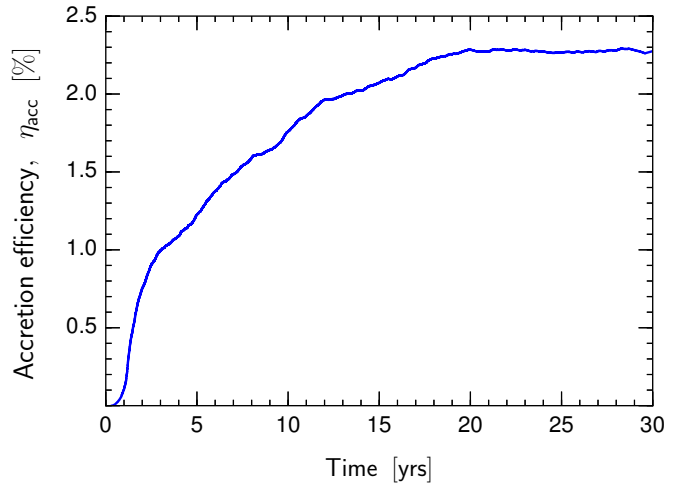}}
  \caption{Mass accretion efficiency as a function of time in the simulation of a binary system consisting of a $3.0\,M_{\sun}$ donor star and a $1.5\,M_{\sun}$ accreting star. The system has an orbital separation of $3\,\rm{au}$, the donor star is losing mass at a rate of $10^{-6}\,M_{\sun}\,\mathrm{yr^{-1}}$ with a wind velocity of $15\,\rm{km\,s^{-1}}$.}
\label{Fig:efficiency}
  \end{center}
\end{figure}

\begin{figure*}
  \begin{center}
    {\includegraphics[width=0.48\textwidth, angle=360]{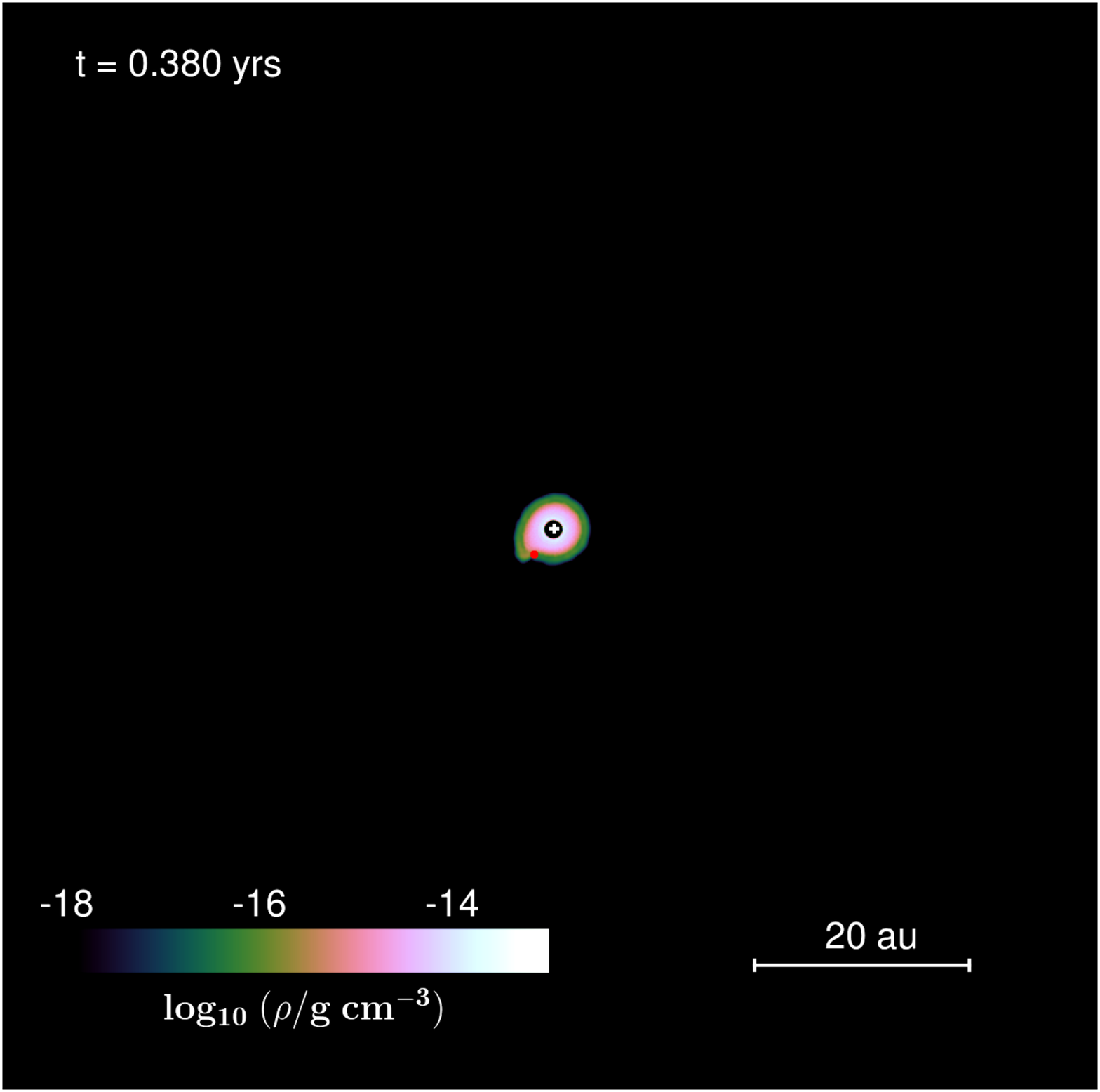}}
    {\includegraphics[width=0.48\textwidth, angle=360]{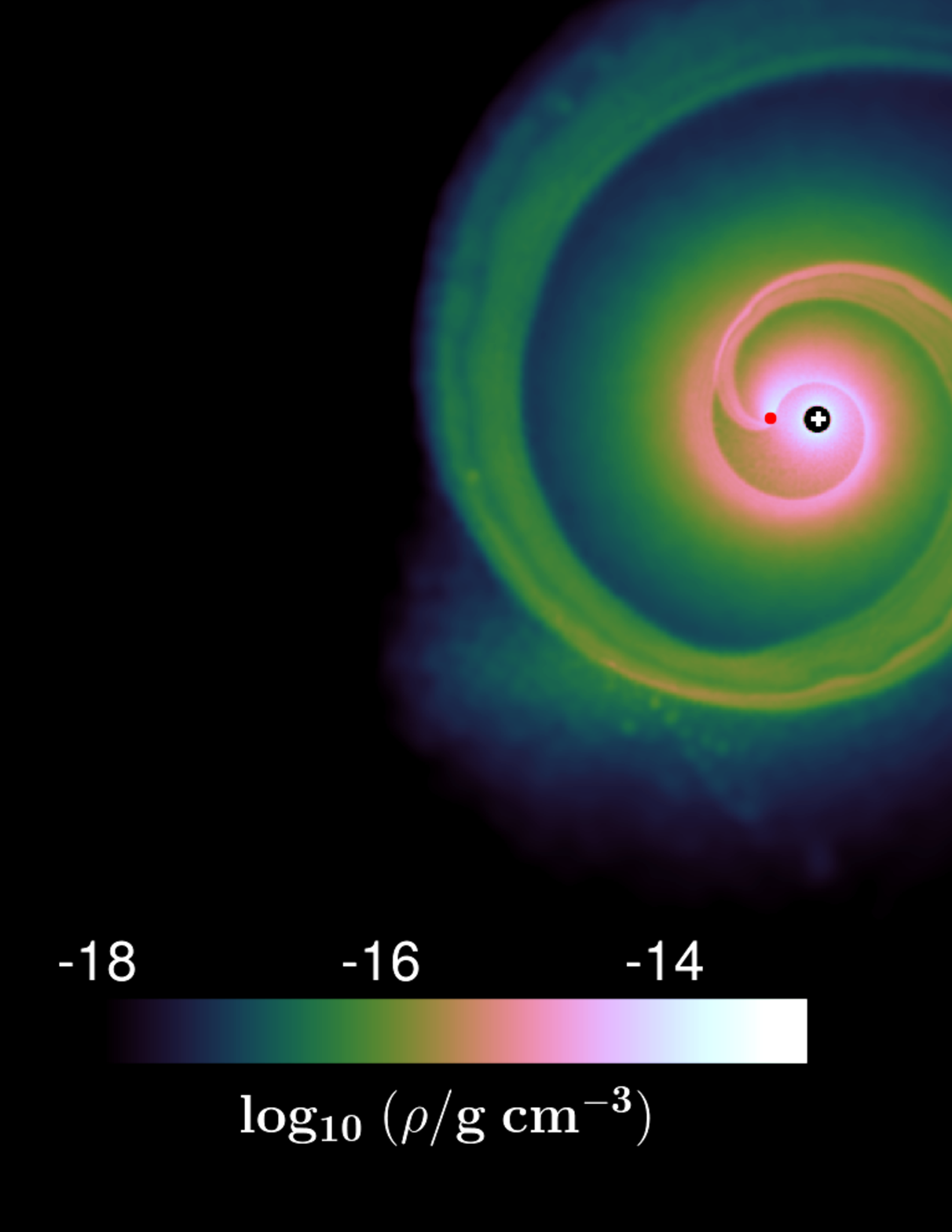}}
    {\includegraphics[width=0.48\textwidth, angle=360]{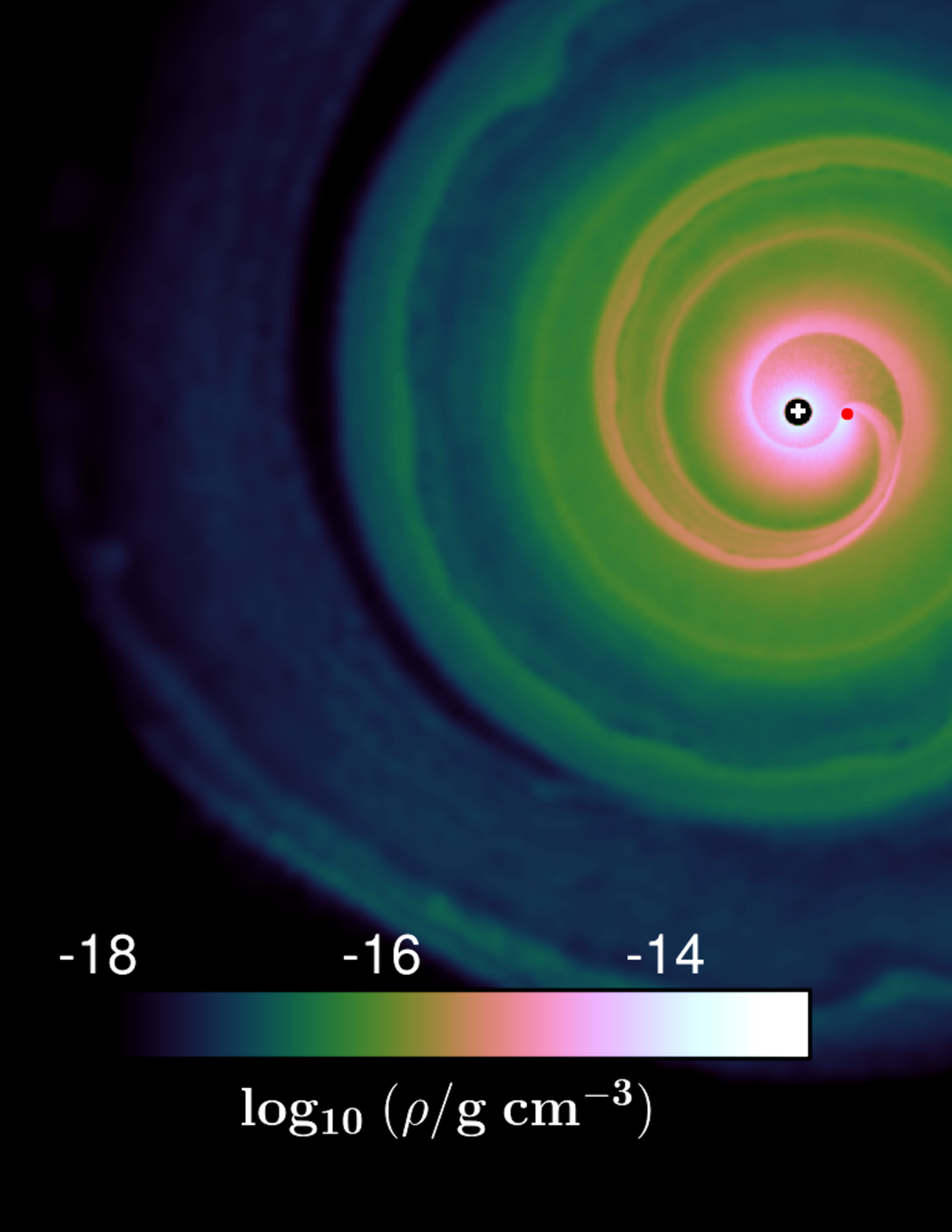}}
    {\includegraphics[width=0.48\textwidth, angle=360]{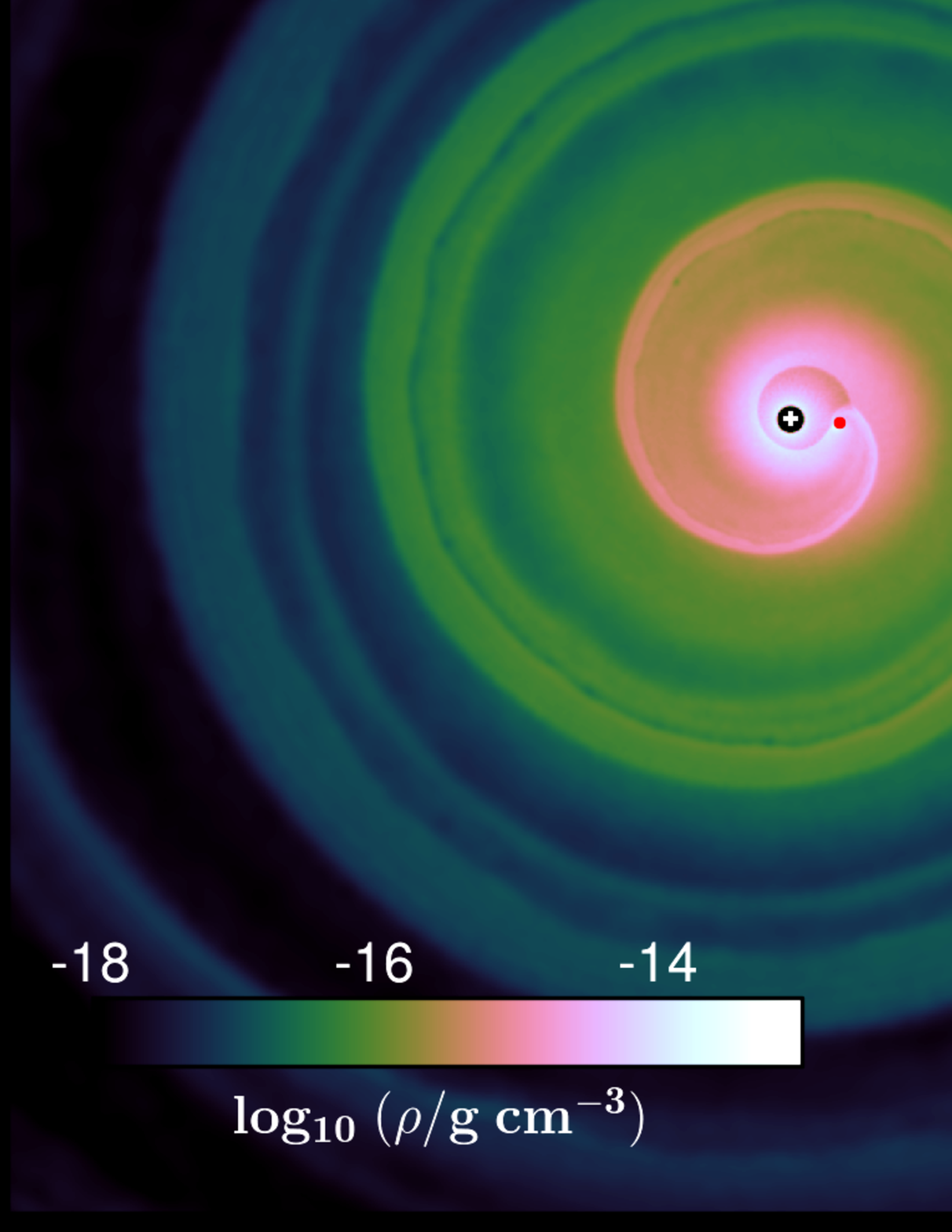}}
  \caption{As Fig.~\ref{Fig:xy1}, but for the binary system with a mass ratio of $q=0.8/3.0$. All initial binary properties but the mass of the accreting star are kept the same.}
\label{Fig:xy2}
  \end{center}
\end{figure*}

\begin{figure*}
  \begin{center}
    {\includegraphics[width=0.32\textwidth, angle=360]{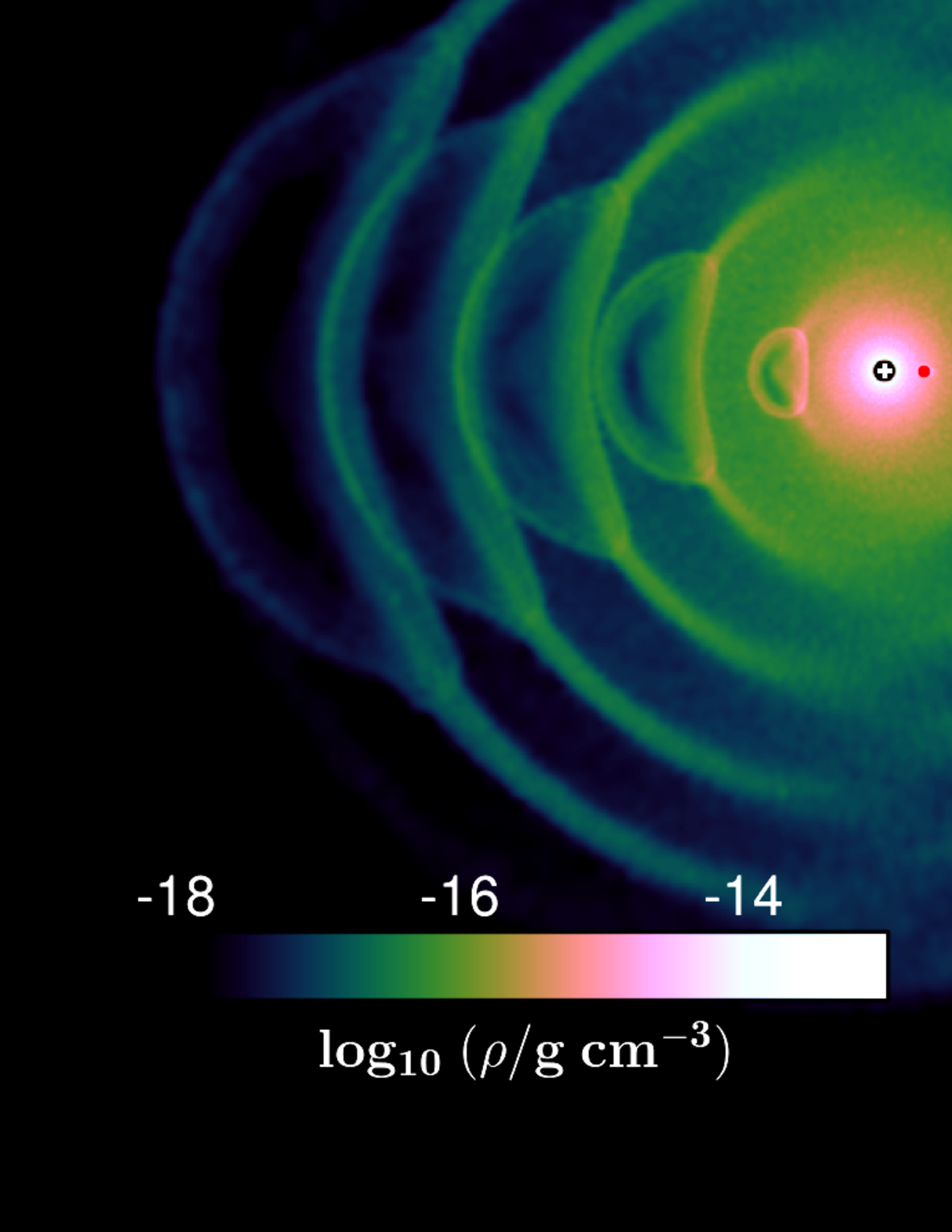}}
    {\includegraphics[width=0.32\textwidth, angle=360]{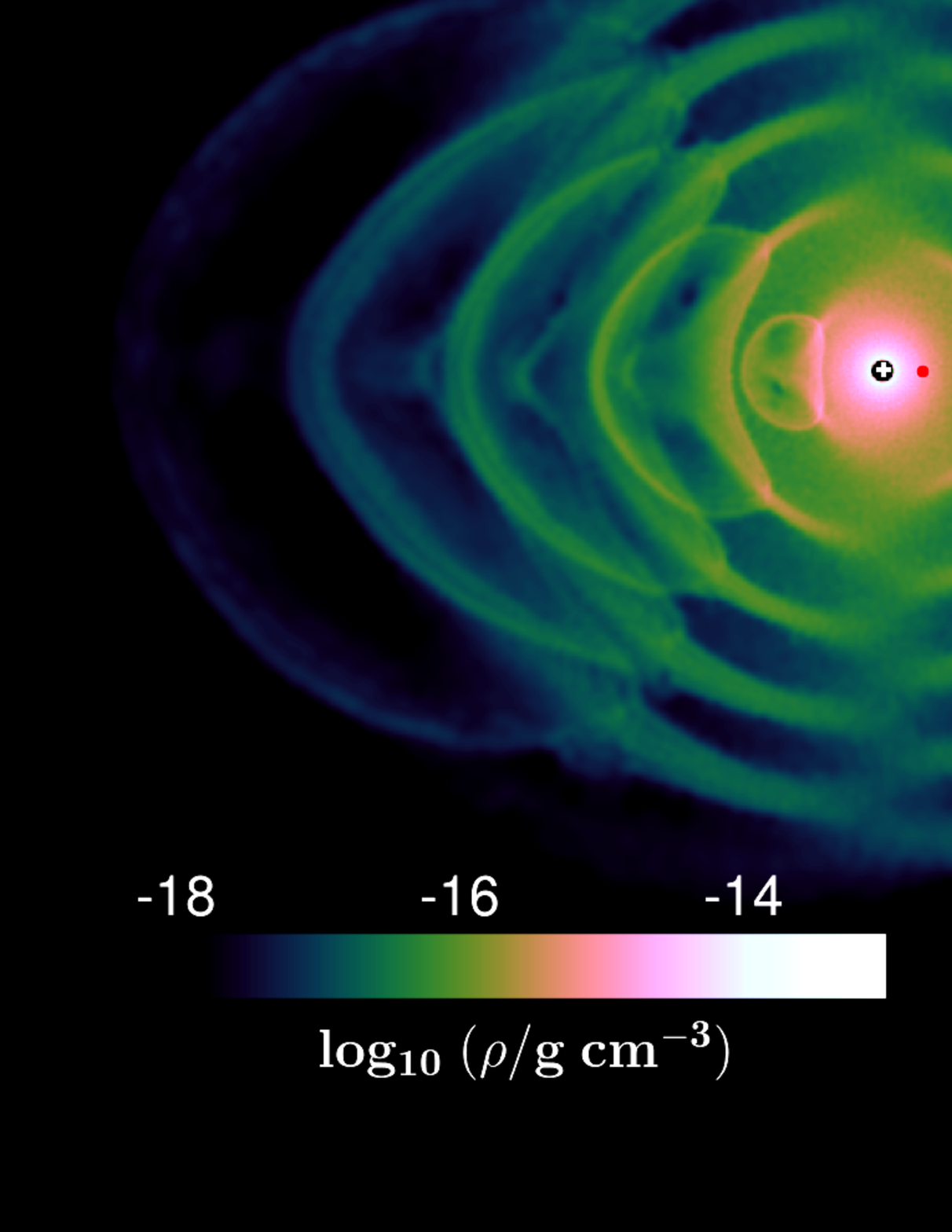}}
    {\includegraphics[width=0.32\textwidth, angle=360]{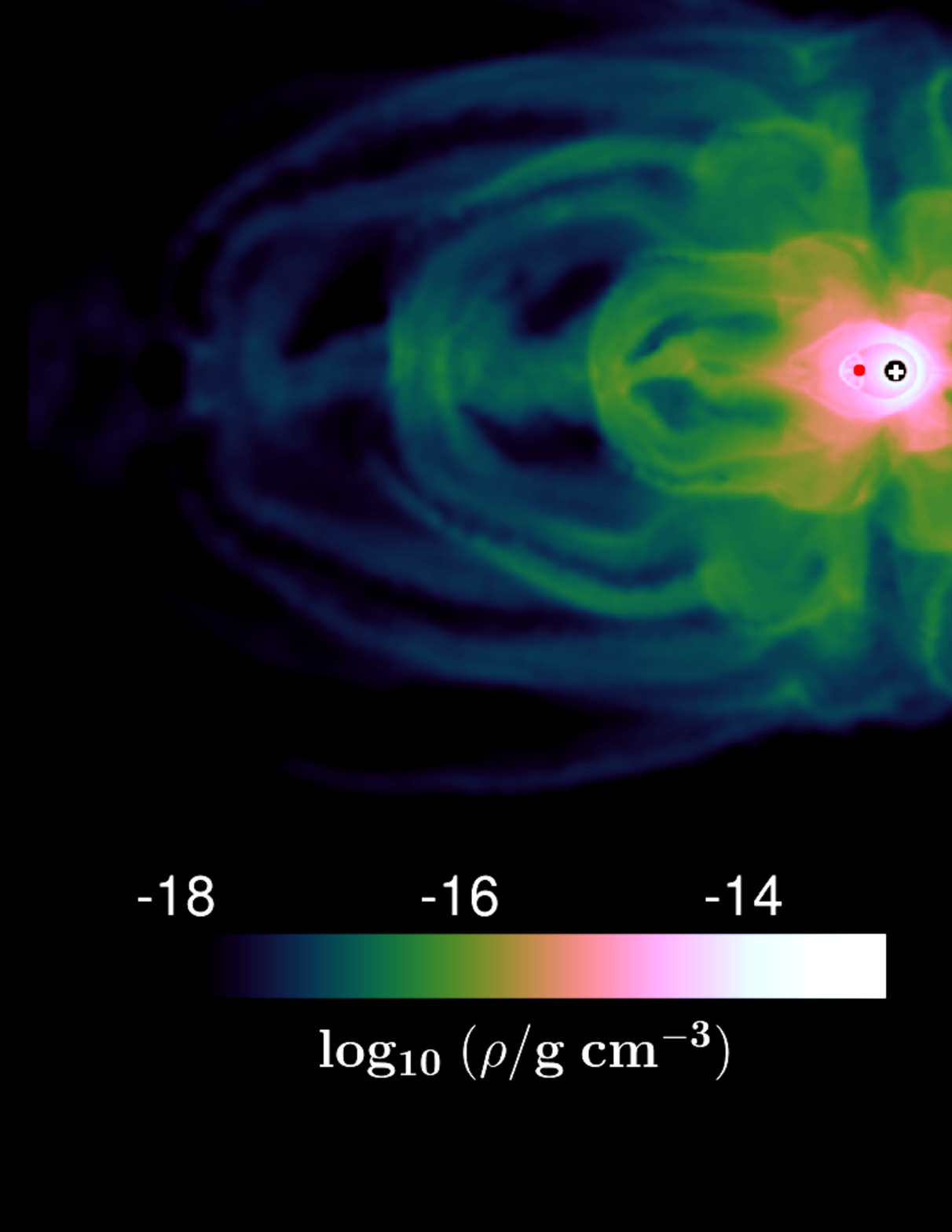}}
    {\includegraphics[width=0.32\textwidth, angle=360]{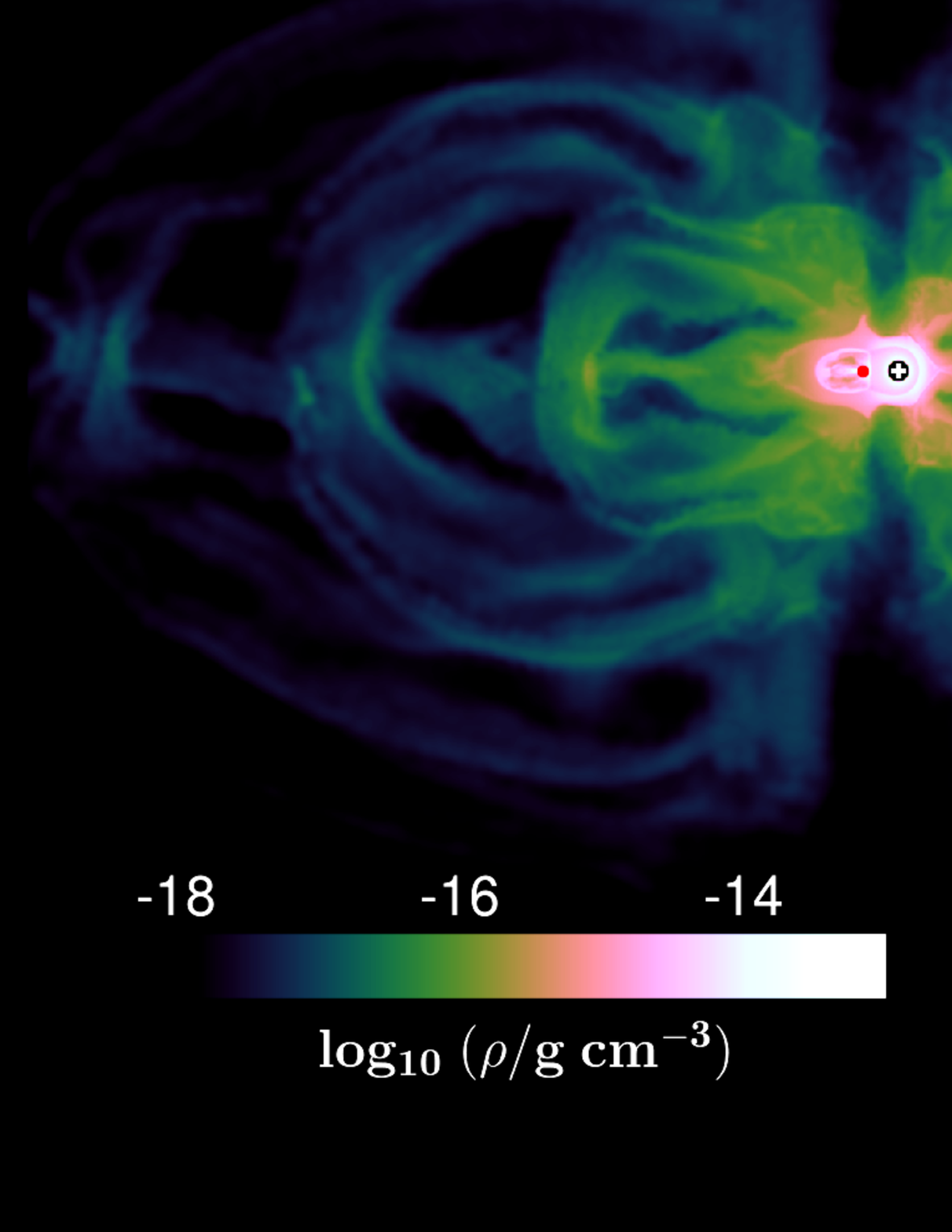}}
    {\includegraphics[width=0.32\textwidth, angle=360]{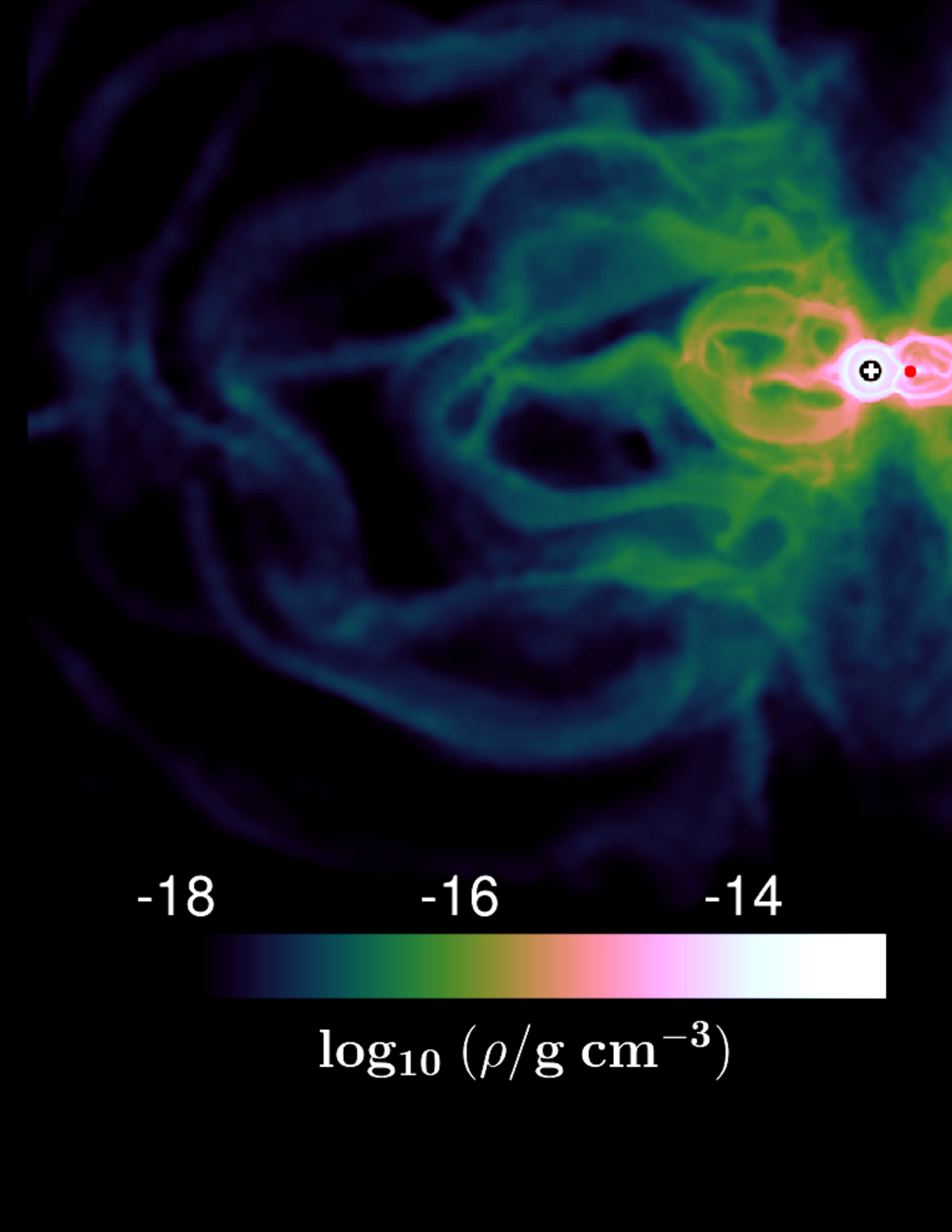}}
    {\includegraphics[width=0.32\textwidth, angle=360]{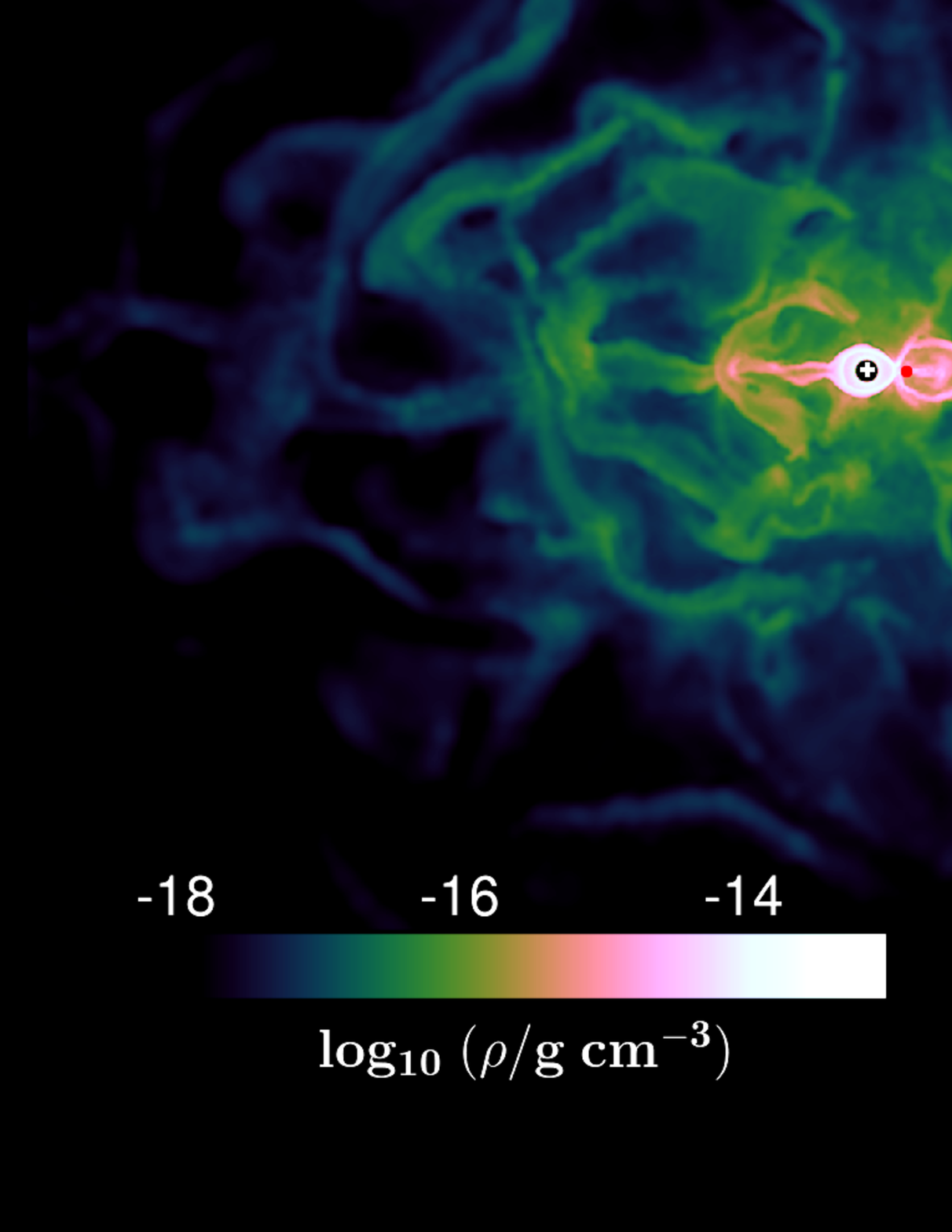}}
  \caption{As Fig.~\ref{Fig:xy1}, but for the density distributions in the $x-z$ plane of binary systems with mass ratios of $q=0.05, 0.27, 0.33, 0.50, 0.67, 1.0$.  
}
\label{Fig:xz}
  \end{center}
\end{figure*}

\begin{figure*}
  \begin{center}
    {\includegraphics[width=0.32\textwidth, angle=360]{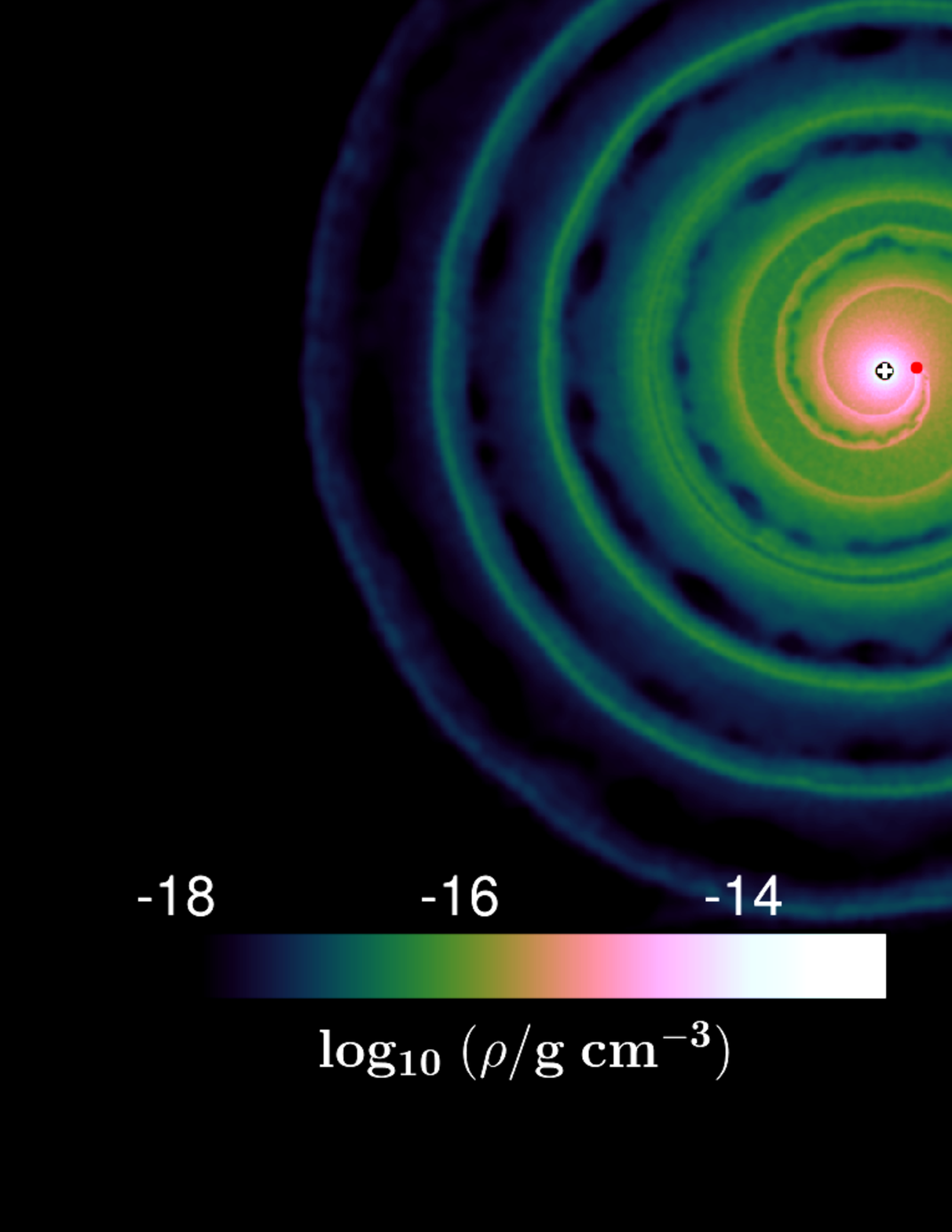}}
    {\includegraphics[width=0.32\textwidth, angle=360]{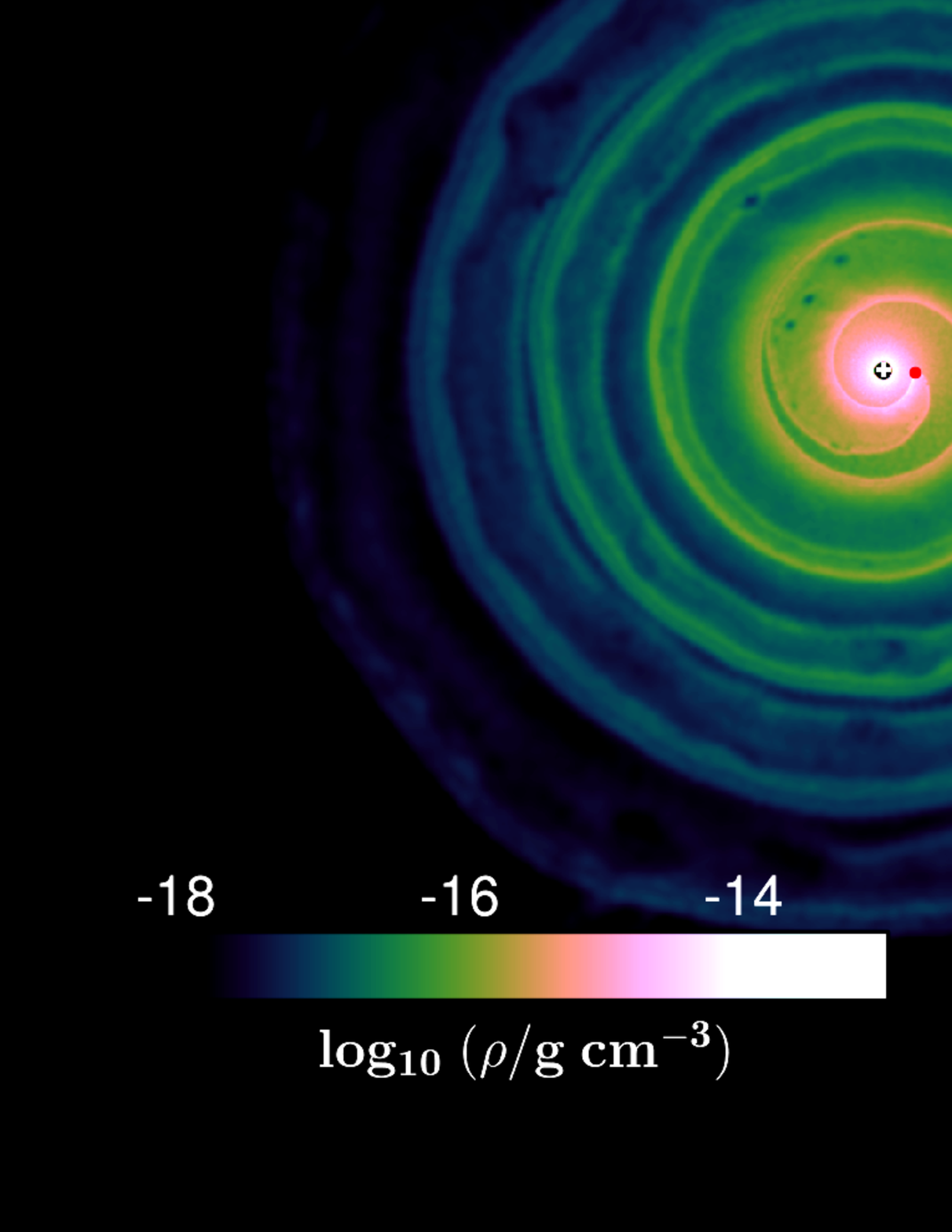}}
    {\includegraphics[width=0.32\textwidth, angle=360]{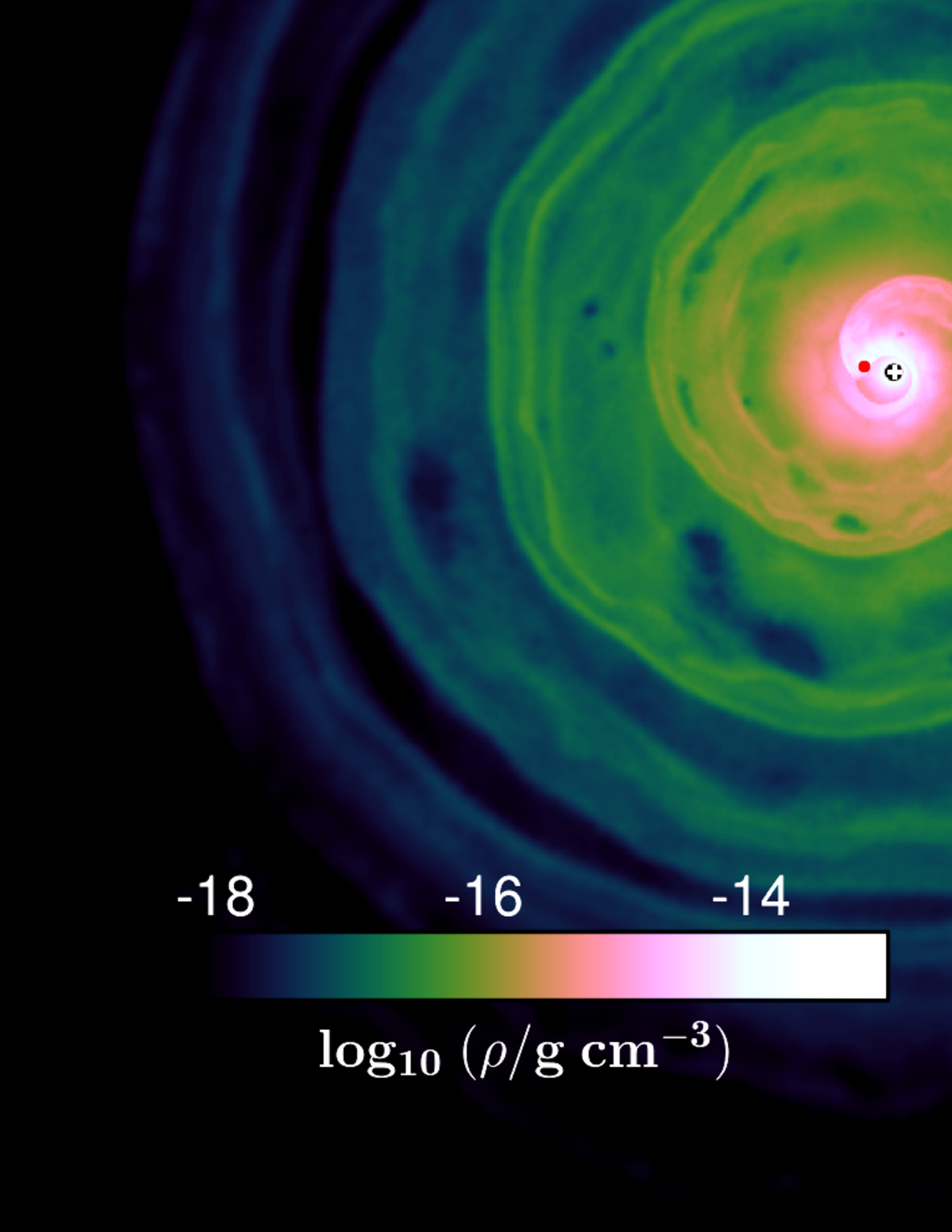}}
    {\includegraphics[width=0.32\textwidth, angle=360]{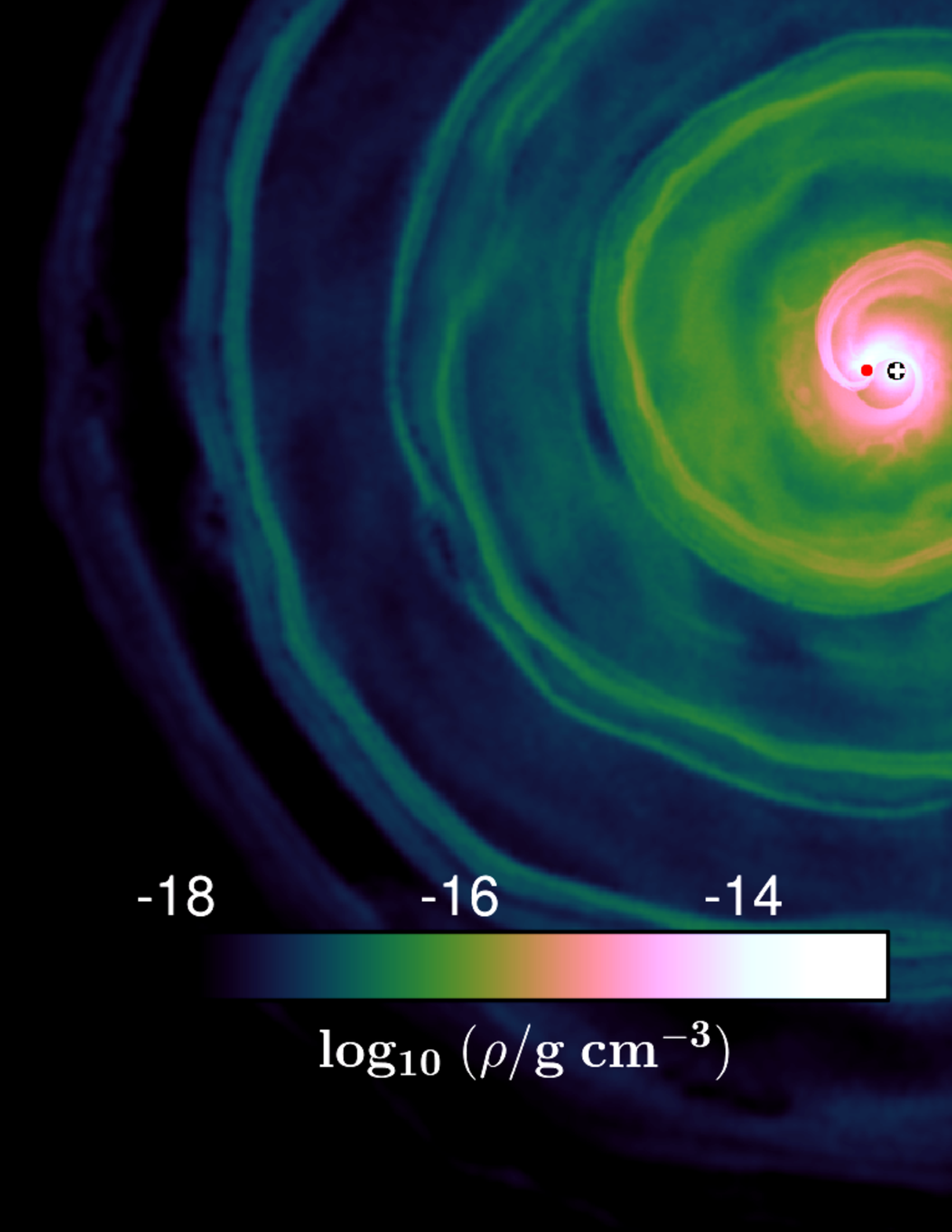}}
    {\includegraphics[width=0.32\textwidth, angle=360]{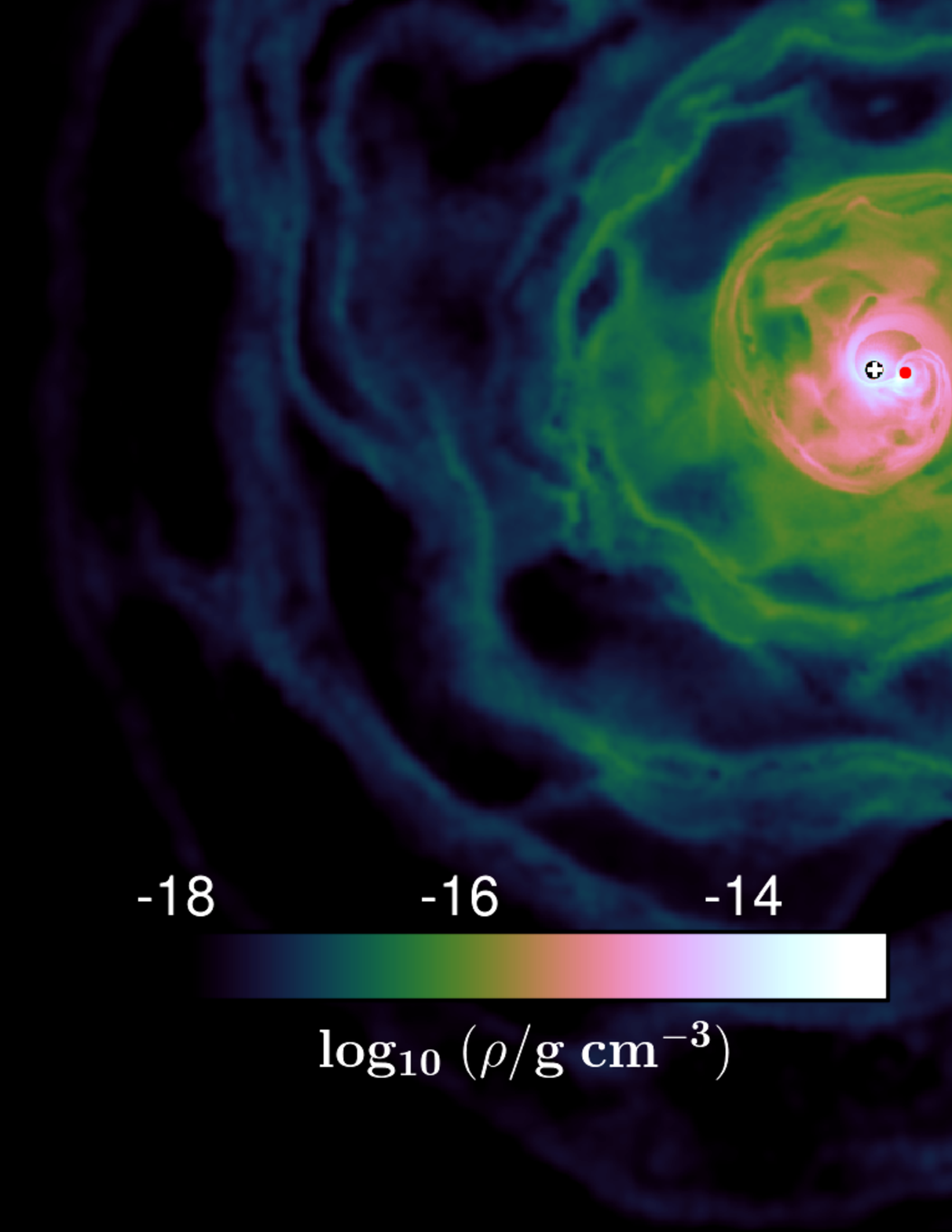}}
    {\includegraphics[width=0.32\textwidth, angle=360]{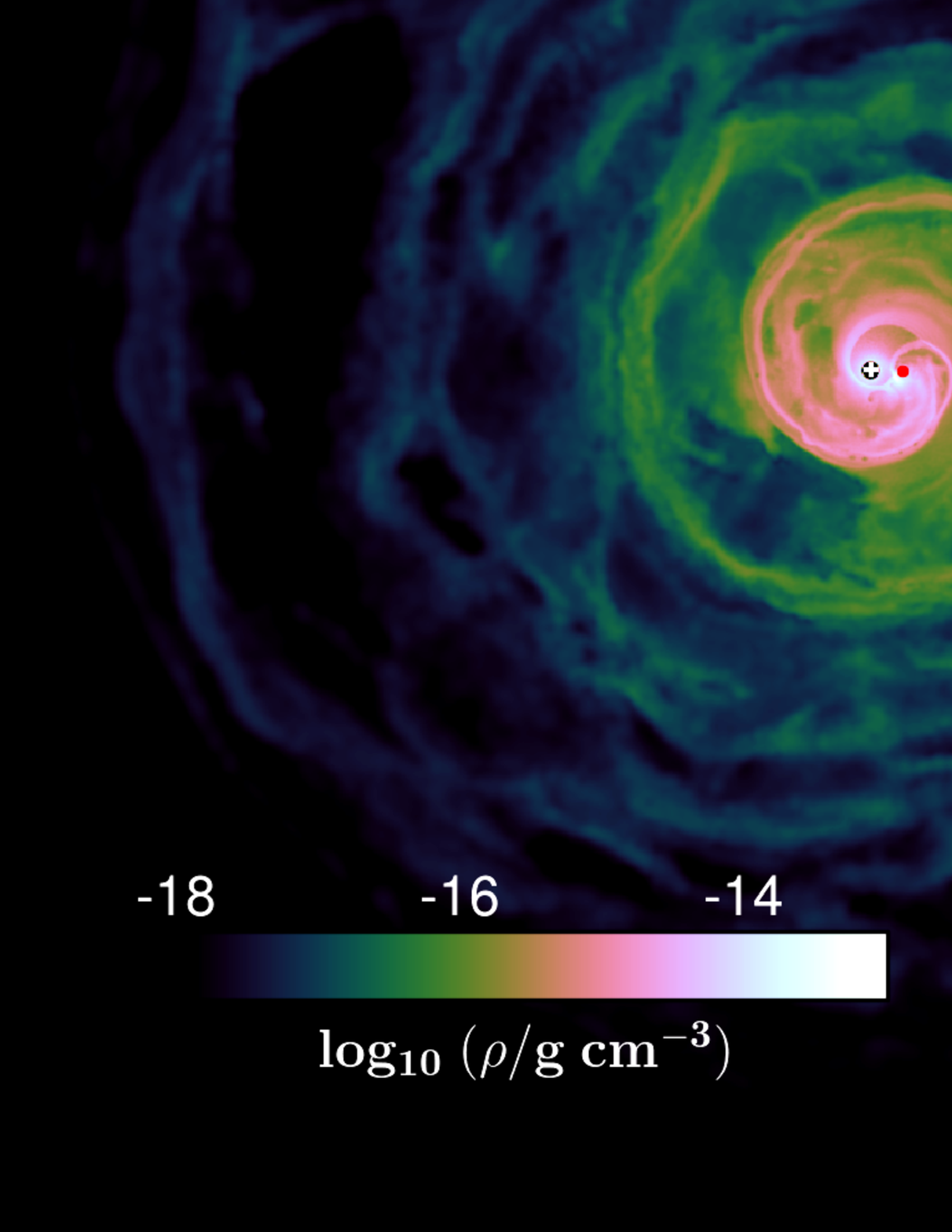}}
  \caption{As Fig.~\ref{Fig:xz}, but for the density distributions in the $x-y$ plane.  
}
\label{Fig:xy3}
  \end{center}
\end{figure*}

We assume the wind is an ideal gas that is characterized by the ratio of specific heats, $\gamma$, in our simulations. A polytropic index in the equation of state for an ideal gas is set to be $\gamma$=$5/3$ in most of the simulations unless specified otherwise. The initial temperature of the gas is $T_{\rm{gas}}\approx 4050/\gamma\ \rm{(K)}$, following \citet{Theuns1996}. A mechanism of wind acceleration due to radiation pressure on dust can be mimicked in the {\sc GADGET} code by implementing an additional force in the momentum equation for a mass element in the wind to reduce the effect of the gravity of the donor star \citep{Theuns1993, Springel2005}:
 \begin{equation}
    \label{eq:1}
     \upsilon\,\frac{\mathrm{d}\,\upsilon}{\mathrm{d}\,r} = \frac{1}{\rho}\frac{\mathrm{d}\,P}{\mathrm{d}\,r}-\frac{G\,M_1}{r^{2}}(1-f),
   \end{equation}
where $f$ varies between 0 and 1 and quantifies the effect of this additional force.

Some studies suggest that the wind of AGB stars is accelerated by a combination of pulsations, which are responsible for dust production at some distance from the stellar surface, and radiation pressure of the photons coming from the AGB star on the dust grains. The dust grains transfer part of their kinetic energy to the gas through collisions and consequently the wind is accelerated outwards \citep{Wood1979,Hill1979,Bowen1988, Theuns1993, Hoefner2003, Woitke2006, Mohamed2010}. However, as the theory of wind acceleration in AGB stars is not well understood, the details of the wind acceleration mechanism are not considered in our present model. For simplicity, we assume that the gravity of the AGB star is balanced by radiation pressure on the dust, i.e., we set $f$=$1$ in our runs. This is sometimes called the ``free wind'' case \citep{Theuns1993, Mohamed2007}.  We assume a constant wind velocity of $\upsilon_{\mathrm{wind}}$=$15\,\mathrm{km\,s^{-1}}$ in all our simulations. 

We do not include the effect of tidal distortions of the mass-losing star which may be important when the primary fills a significant portion of its Roche lobe. Also, effects due to rotation, magnetic fields, and radiation of the accreting star are not included.

\subsection{Mass Accretion onto the Secondary}
\label{sec:accretion}

Ideally, we think a particle is accreted by the secondary star once it reaches its surface. However, the accreting star in our simulations is a low-mass main sequence star (or a white dwarf) with a stellar radius within a few solar radii, which is more than two orders of magnitude smaller than the typical orbital separation (a few $\rm{au}$) adopted in our wind-accreting binary systems. Because of the Lagrangian nature of SPH, it is computationally very challenging to accurately resolve the surface of the accreting star in the simulation. As more and more particles move to the higher density region, the time-steps of the simulations will progressively decrease. Consequently, the calculations will soon become too slow to be feasible. Two methods are generally adopted for simulating mass accretion by the secondary in wide binaries to overcome this problem: the so-called ``prompt accretion'' and ``gradual accretion''.  In the former, wind particles that come within an accretion radius $r_{\rm{acc}}$  of the accreting star are promptly removed from the calculation and assumed to be accreted by the accreting star. In the latter, a point ``sink'' term is included into the continuity equation \citep{Anzer1987, Mastrodemos1998}, and a virtual mass $m_i$ is assigned to each particle $i$ within a certain distance $r_{\rm{h}}$ from the surface of the secondary star. This virtual mass varies as a function of the distance between the particle and the secondary star, $r_{i\,2}$. The variation of $m_i$ is calculated as follows:
 \begin{equation}
    \label{eq:2}
     \frac{\mathrm{d}\,m_{i}}{\mathrm{d}\,t} = \mathrm{min} \left [ -\frac{m_{i}\,(1-r_{i\,2}^{2}/\bar{h}^{\,2})}{\mathrm{d}\,t},\ 0\,\right ],
   \end{equation}
where $\mathrm{d}m_{i}$ is the change in mass of particle $i$ with an original mass of $m_{i}$, $\mathrm{d}\,t$ is the time-step, $\bar{h}$ is the average smoothing length of all SPH particles within $r_{\rm{h}}$, and $r_{i\,2}$ is the distance of particle $i$ to the accreting star. As the virtual mass of the particle reduces to be a small fraction (typically a few $0.1\%$) of the original mass, this particle is assumed to be accreted \citep{Anzer1987}.

In this study, following \citet{Theuns1993} and \citet{Mohamed2012}, the ``gradual accretion'' is adopted to simulate mass accretion onto the secondary. Here, $r_{\mathrm{h}}$ is chosen to be the order of the local average smoothing length, i.e., $r_{\mathrm{h}}\approx 0.5\,\mathrm{au}$. Wind particles are removed from the calculation and assumed to be accreted by the secondary when their virtual mass is reduced to $0.1\%$ of the original mass. Comparing with the prompt accretion, this method is more numerically stable at the accretion boundary and ensures that no pressure builds up around the accreting star.  For a detailed discussion of accreting particles in SPH simulations, we refer the reader to \citet{Theuns1993} and \citet{Bate1995}.

\section{Results}
\label{sec:results}

\begin{figure*}
  \begin{center}
    {\includegraphics[width=0.32\textwidth, angle=360]{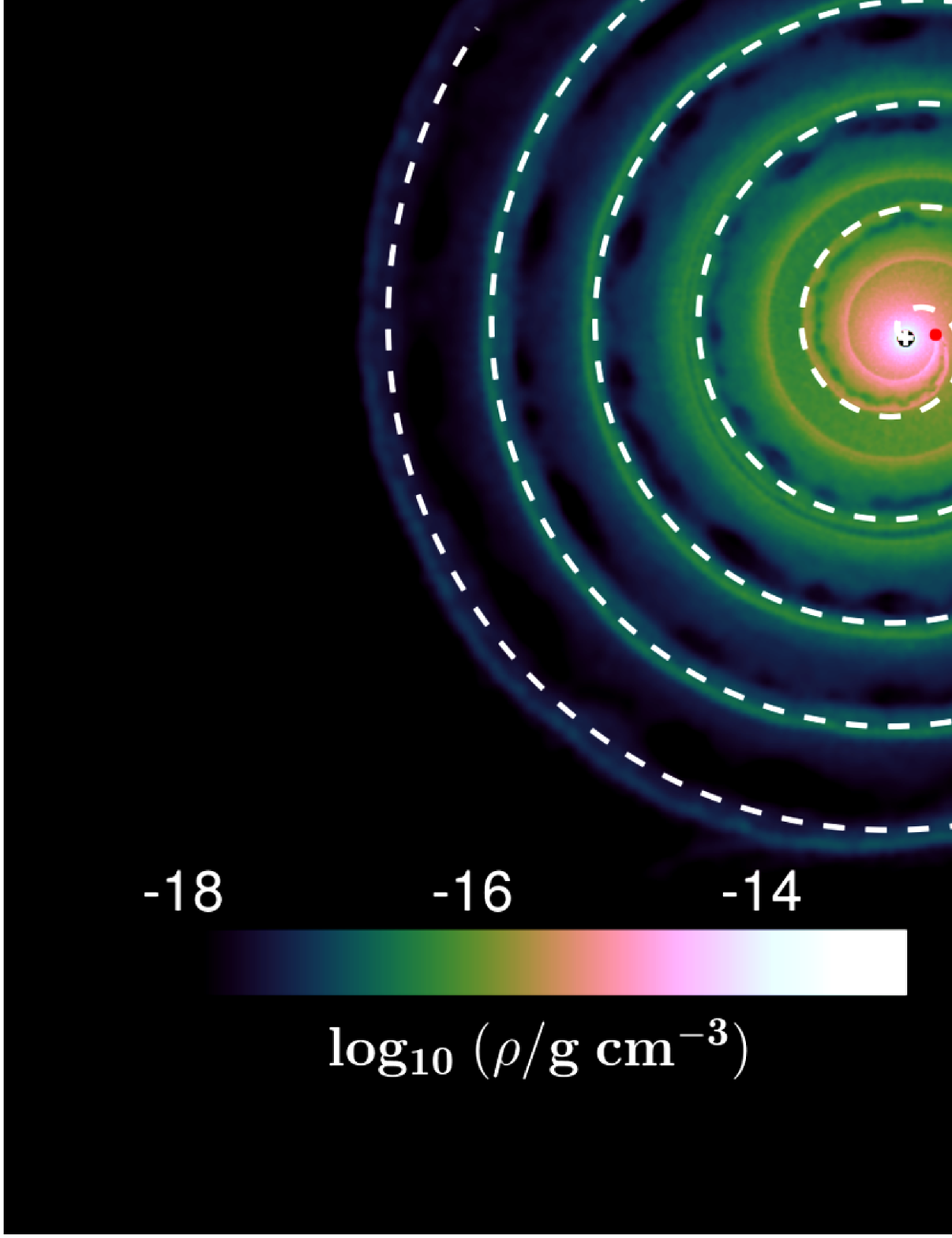}}
    {\includegraphics[width=0.32\textwidth, angle=360]{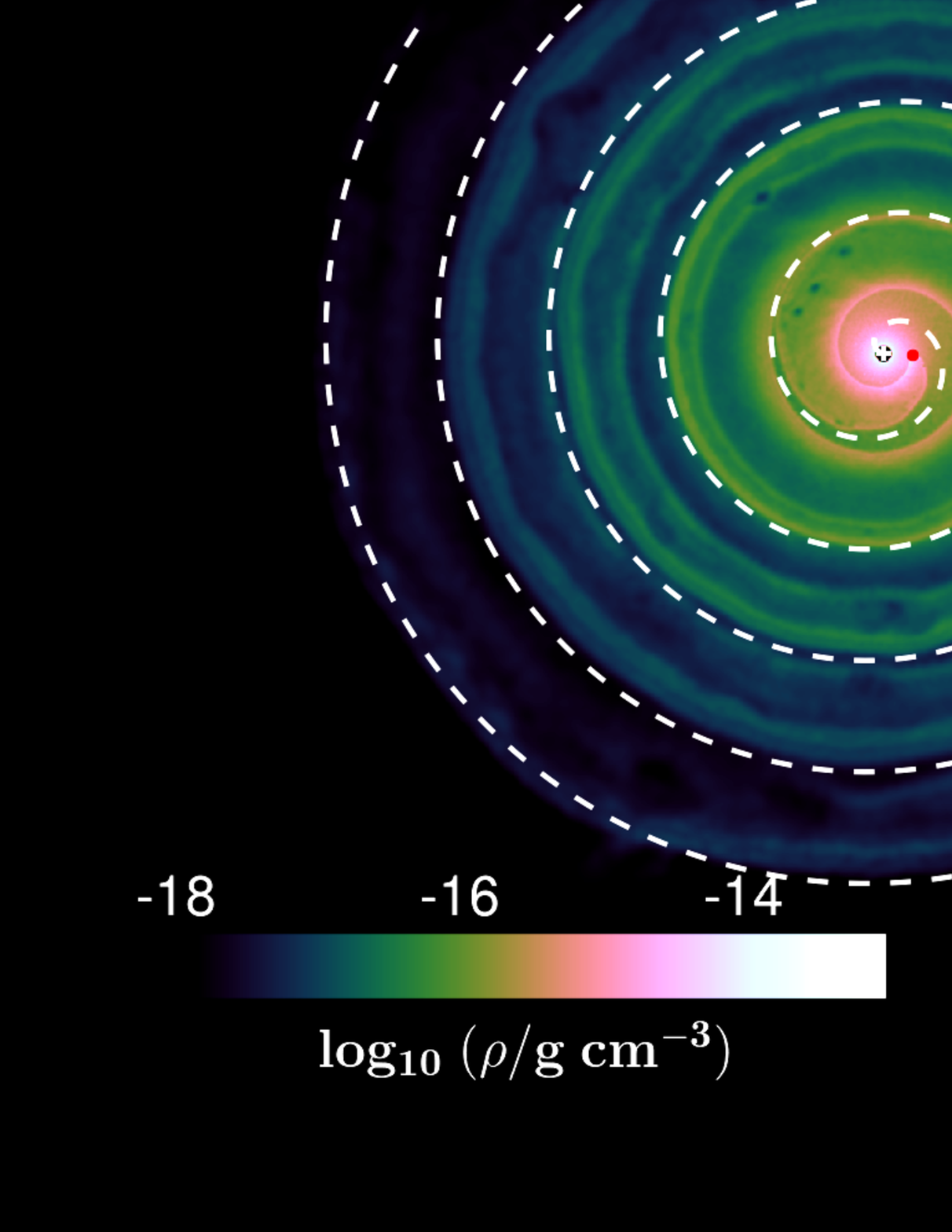}}
    {\includegraphics[width=0.32\textwidth, angle=360]{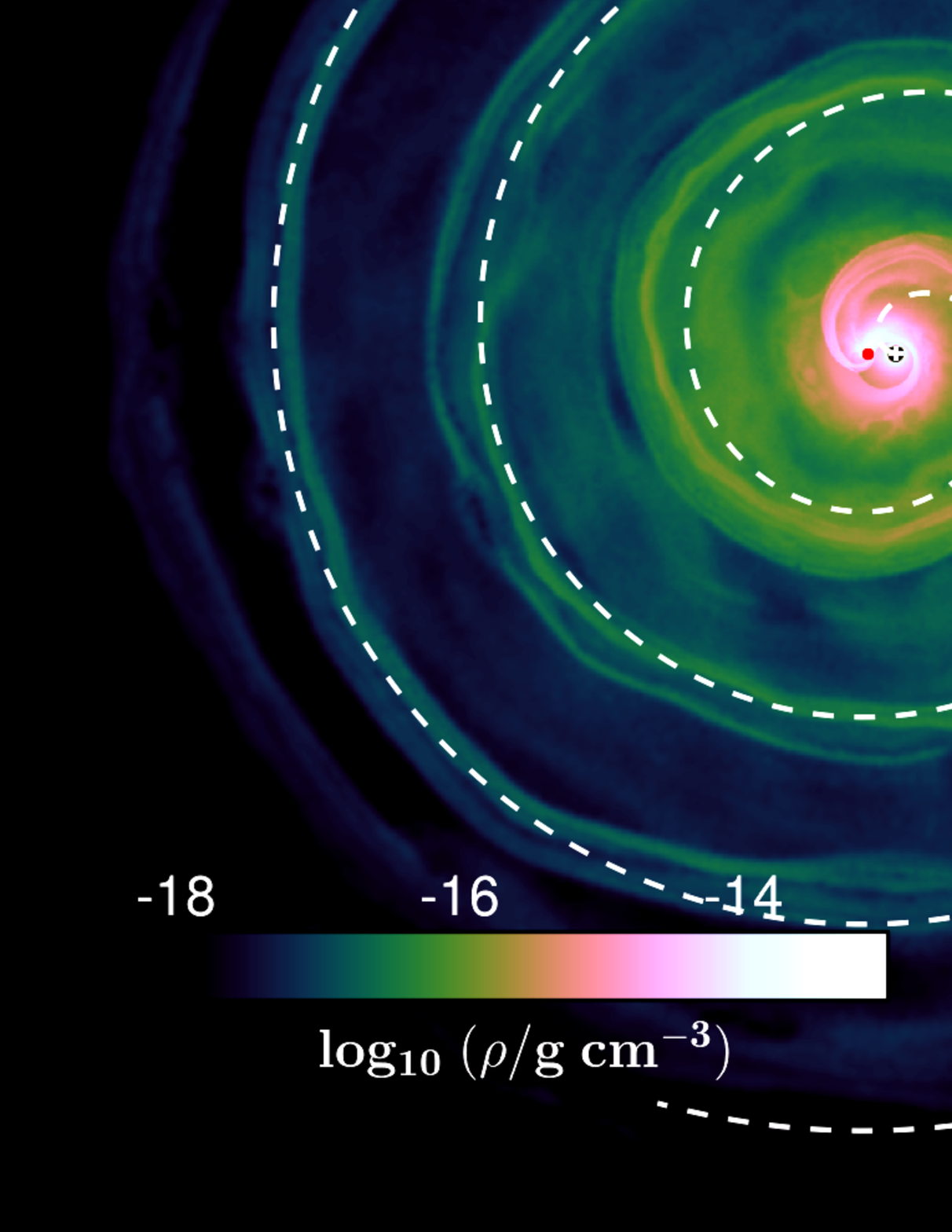}}
  \caption{As Fig.~\ref{Fig:xy3}, but for the $x-y$ plane of the simulations with model M01, M02 and M08 model, respectively (Table~\ref{Tab:1}). The spiral pattern visible in the $x-y$ plane has been fitted with a spiral function, $r$=$b\theta$ (white-dashed curves). The constant $b$ is a free parameter of the fit and it is about $2.55\times10^{13}\,\mathrm{cm}$, $2.80\times10^{13}\,\mathrm{cm}$ and $5.21\times10^{13}\,\mathrm{cm}$ for the mass ratio of $0.05$ (left-hand panel), $0.1$ (middle panel) and $0.5$ (right-hand panel), respectively.}
\label{Fig:fitxy}
  \end{center}
\end{figure*}

In this section we present the results of our simulations of binary systems consisting of a primary with a fixed mass of $M_1$=$3.0\,M_{\odot}$ and a secondary with a mass in the range between $0.15\,M_{\odot}$ and $3.0\,M_{\odot}$, corresponding to mass ratios $q$=$M_2/M_1$ between $0.05$ (such low mass ratio may play an important role in the formation and shaping of PNe \citealt{deMarco2011}) and $1.0$. A fixed orbital separation $A_{\mathrm{orb}}$=$3.0\,\mathrm{au}$ is used in all simulations. This setup allows us to study the effect of different mass ratios on the results. A summary of the secondary masses, mass ratios, orbital velocities, and the mass and angular momentum transferred in each model is presented in Table~\ref{Tab:1}. All values of $\eta_{\mathrm{acc}}$ and $j_{\mathrm{acc}}$ presented in this work are given once a relatively steady state has been reached, i.e., all simulations are evolved for a few tens of orbital periods (Fig.~\ref{Fig:efficiency}).

\begin{table} \renewcommand{\arraystretch}{1.0}
\fontsize{8}{11}\selectfont
\begin{center}
\caption{The results of our simulations of binary systems with different mass ratios. \label{Tab:1}}
\begin{tabular*}{.49\textwidth}{@{\extracolsep{\fill}}lcccclc}
\tableline\tableline
Model & $M_{\mathrm{2}}$ & $q$  & $v_{\mathrm{orb1}}$     & $v_{\mathrm{orb2}}$     &\  $\eta_{\mathrm{acc}}$ & $j_{\mathrm{acc}}$\\
      &  ($M_{\sun}$)    &     & $(\mathrm{km/s})$ & $(\mathrm{km/s})$ &\ (\%)                   & $\mathrm{(10^{18}cm^{2}/s)}$\\
\tableline
M01 & 0.15 & 0.05 & 1.5  & 29.1 &\  0.12 & 1.17\\
M02 & 0.30 & 0.10 & 2.8  & 28.4 &\  0.40 & 1.25\\
M03 & 0.60 & 0.20 & 5.4  & 27.2 &\  0.96 & 1.42\\
M04 & 0.80 & 0.27 & 7.1  & 26.5 &\  1.47 & 1.51\\
M05 & 0.90 & 0.30 & 7.8  & 26.1 &\  1.77 & 1.63\\
M06 & 1.00 & 0.33 & 8.6  & 25.8 &\  1.98 & 1.68\\
M07 & 1.20 & 0.40 & 10.1 & 25.2 &\  2.16 & 1.77\\
M08 & 1.50 & 0.50 & 12.2 & 24.3 &\  2.28 & 1.81\\
M09 & 1.80 & 0.60 & 14.1 & 23.6 &\  2.91 & 1.90\\
M10 & 2.00 & 0.67 & 15.4 & 23.1 &\  3.72 & 1.98\\
M11 & 2.40 & 0.80 & 17.8 & 22.2 &\  5.88 & 2.23\\
M12 & 2.60 & 0.87 & 18.9 & 21.8 &\  7.20 & 2.32\\
M13 & 2.80 & 0.93 & 20.0 & 21.4 &\  7.38 & 2.35\\
M14 & 3.00 & 1.00 & 21.1 & 21.1 &\  7.56 & 2.38\\

\tableline
\end{tabular*}
\end{center}

\medskip
Note: Here, $M_{2}$ is the mass of the accreting companion star; $q$=$M_{2}/M_{1}$ is the mass ratio of the binary system; $v_{\mathrm{orb1}}$ and $v_{\mathrm{orb2}}$ are initial orbital velocity of the mass-losing star and accreting companion star, respectively; $\eta_{\mathrm{acc}}$ is the mass accretion efficiency; $j_{\mathrm{acc}}$ is the average accreted specific angular momentum. The mass of the mass-losing star $M_{1}$=$3.0\,M_{\sun}$, the radius of the mass-losing star $R_{1}$=$200\,R_{\sun}$, the orbital separation of the systems $A_{\mathrm{orb}}$=$\,3.0\,\mathrm{au}$, the mass-loss rate $\dot{M}_{\mathrm{wind}}$=$\,10^{-6}\,M_{\sun}\,\mathrm{yr^{-1}}$, and the wind speed of $\upsilon_{\mathrm{wind}}$=$\,15\,\mathrm{km\,s^{-1}}$ are fixed in all simulations.\\

\end{table}

\subsection{A Description of Wind Mass Transfer}
\label{sec:description}

Figure~\ref{Fig:xy1} shows the evolution of the gas density in the orbital plane ($x-y$) as a function of time in the simulation of the model M07 (Table~\ref{Tab:1}). The M07 model is chosen as our default model in this work, its binary properties and wind parameters are set to be consistent with those adopted by \citet{Theuns1996} in their model with $\gamma$=$1.5$. Here, the donor star is losing mass at a rate of $10^{-6}\,\rm{M_{\sun}\,yr^{-1}}$ with a wind velocity of $15\,\rm{km\,s^{-1}}$. In general, a vortex structure is seen around the accreting companion. As time passes and more gas flows beyond the accreting companion star (panels~2--4), the spiral-shaped accretion stream forms, and the accreting star starts to capture material lost by the donor star. The presence of a large spiral is different from the simple, symmetric flow structure in BHL accretion (see also \citealt{Theuns1993}). Also, it is clearly shown that the whole binary system is surrounded by an expanding spiral arm structure as time goes on, and turbulence of the gas within the spiral arms is also seen (see also \citealt{Theuns1993, de-Val-Borro2009, Toupin2015}).

\begin{figure*}
  \begin{center}
    {\includegraphics[width=0.45\textwidth, angle=360]{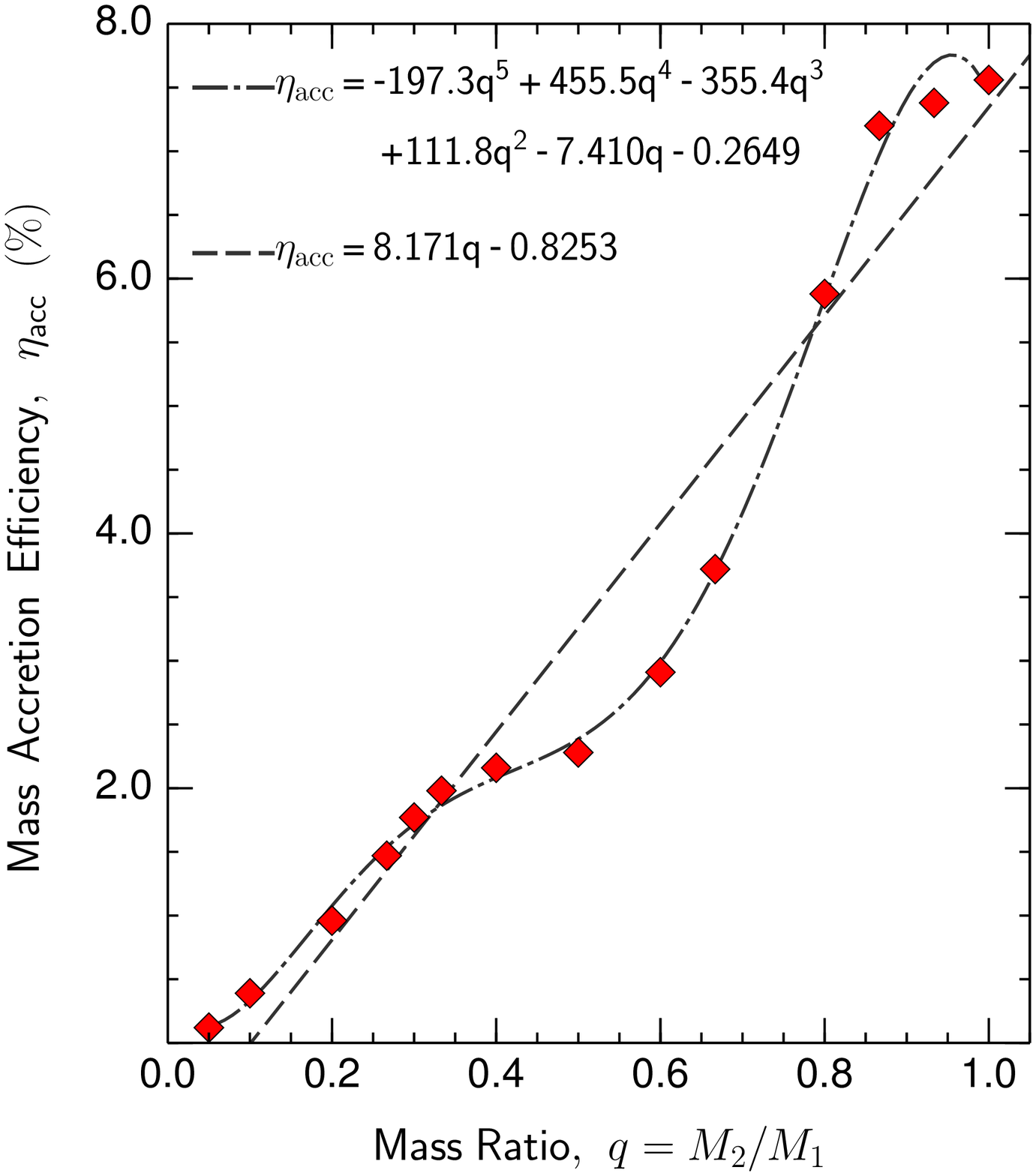}}
    \hspace{0.45in}
    {\includegraphics[width=0.45\textwidth, angle=360]{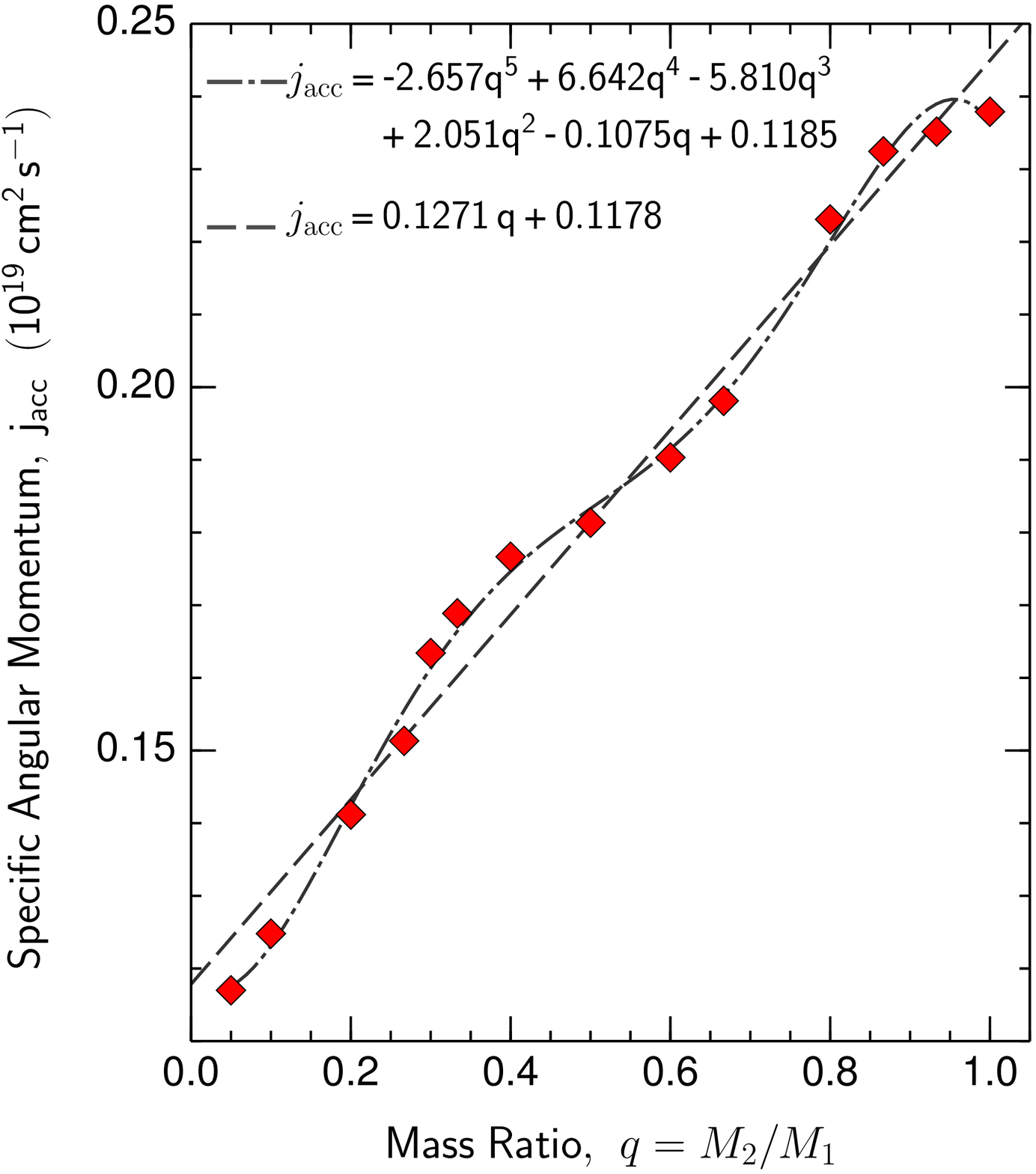}}
  \caption{Mass accretion efficiency (left panel) and average accreted specific angular momentum (right panel) as a function of the mass ratio ($M_2/M_1$) of the binary system. Diamond markers are the results from our simulations. Empirical fifth order polynomial and linear fits are shown by dash-dotted and dashed lines, respectively. Fit parameters are also given in figures.}
\label{Fig:fit}
  \end{center}
\end{figure*}

Figure~\ref{Fig:efficiency} gives the mass accretion efficiency, $\eta_{\mathrm{acc}}$, as a function of time. Here, we calculate the mass accretion efficiency as the ratio of mass accreted by the secondary ($\dot{M}_{\mathrm{acc}}$) to the mass-loss rate ($\dot{M}_{\mathrm{wind}}$) from the primary averaged over 100 time-steps (which roughly corresponds to a half of year). The mass accretion efficiency initially increases with time as the wind particles are injected into the simulation around the donor star. It eventually reaches a relatively stable value of $\eta_{\mathrm{acc}}\approx2.3\%$ after about 12 orbital periods, which is about ten times smaller than theoretical estimates based on the BHL prescription. This is consistent with the results of \citet{Theuns1996} in their simulation for a system with the same initial parameters. In addition, we find that the specific angular momentum accreted by the secondary is $j_{\mathrm{acc}}\approx1.8\times10^{18}\,\mathrm{cm^{2}\,s^{-1}}$, which is one order of magnitude lower than the Keplerian value at an accretion radius ($R_{\mathrm{a}}$=$2\,GM_{2}/v^{2}$, \citealt{Hoyle1939}) of about $7\times10^{19}\,\mathrm{cm^{2}\,s^{-1}}$. Here, $j_{\mathrm{acc}}$ is calculated by averaging the specific angular momentum of all accreted particles when they are removed from the simulations. As described in Section~\ref{sec:accretion}, because details of the accreting star are not numerically resolved in our simulations, we cannot trace the particle until it reaches the actual surface of the star. In a realistic case, some physical mechanism such as the magnetic fields may cause an evolution of the specific angular momentum of the wind particle from the place where it is removed to the actual surface of the accreting star, which is neglected in our calculation here.

\subsection{Mass ratio Dependency}
\label{sec:mass_ratio}

We have computed a set of simulations in which we vary the initial mass of the accreting star, $M_{2}$, and keep all other parameters constant to investigate the dependence of the accreted mass and angular momentum on the mass ratio of the binary system, $q$=$M_{2}/M_{1}$. An overview of the models with different mass ratios is given in Table~\ref{Tab:1}. Figure~\ref{Fig:xy2} presents the evolution of gas density in the $x-y$ plane in the simulation for a binary system with a mass ratio $q$=$0.8/3.0$. The basic morphology of the outflow is similar to that in Fig.~\ref{Fig:xy1}. A large sprial of accretion flow is also seen. However, the turbulence of the gas within the spiral arms seen in Fig.~\ref{Fig:xy2} is much stronger than those in Fig.~\ref{Fig:xy1}.

A comparison of the geometries of the outflowing gas in the $x-z$ and $x-y$ plane for simulations with different mass ratios is shown in Figs.~\ref{Fig:xz} and Figs.~\ref{Fig:xy3}, respectively. In general, it is shown that the mass ratio influences details of the global geometry of the gas, as the mass ratio increases the flow structure becomes more and more complicated. The asymmetric features and turbulence of accretion flow and outflowing gas become much more significant in the case with a high mass ratio. Fig.~\ref{Fig:xz} shows that the outflowing gas exhibits a ``multiple-rings'' structure in the M01 model with the lowest mass ratio of 0.05 (Table~\ref{Tab:1}). As the mass ratio increases, the flow structure in the $x-z$ plane tends to flatten and the structure with clearly visible spherical rings is progressively destroyed. The structure of the outflow becomes increasingly complex with higher mass ratios, qualitatively resembling a ``spider-like'' shape ($q=0.33$)  and a ``rose-like'' shape ($q=1.0$). Fig.~\ref{Fig:xy3} shows that a large spiral structure is clearly present in the $x-y$ plane in the simulations with a low mass ratio. However, as the mass ratio increases, the spiral structure becomes less neat and it is visible only in the central region. As shown in Fig.~\ref{Fig:fitxy}, we can roughly fit the structure seen in the simulations by using a spiral function in polar coordinates, $r=b\,\theta$, where $r$ and $\theta$ are the radial and angular coordinate, respectively, and $b$ is a free parameter of the fit. We obtain that the constant $b$ is about $2.55\times10^{13}\,\mathrm{cm}$, $2.80\times10^{13}\,\mathrm{cm}$ and $5.21\times10^{13}\,\mathrm{cm}$ for a mass ratio of $0.05$ (M01 model), $0.1$ (M02 model) and $0.5$ (M08 model), respectively. These correspond to a distance between spirals ($2\pi b$) of around $1.57\times10^{14}\,\mathrm{cm}$, $1.76\times10^{14}\,\mathrm{cm}$ and $3.27\times10^{14}\,\mathrm{cm}$, respectively. The distance between spirals increases by a factor of two as the mass ratio increases by a factor of ten. For $q>0.6$, the spiral pattern of the simulations is rather blurry, except in the central regions, and consequently it is not possible to determine a satisfactory fit of the outflow with a simple form $r=b\,\theta$.

\begin{figure*}
  \begin{center}
    {\includegraphics[width=0.495\textwidth, angle=360]{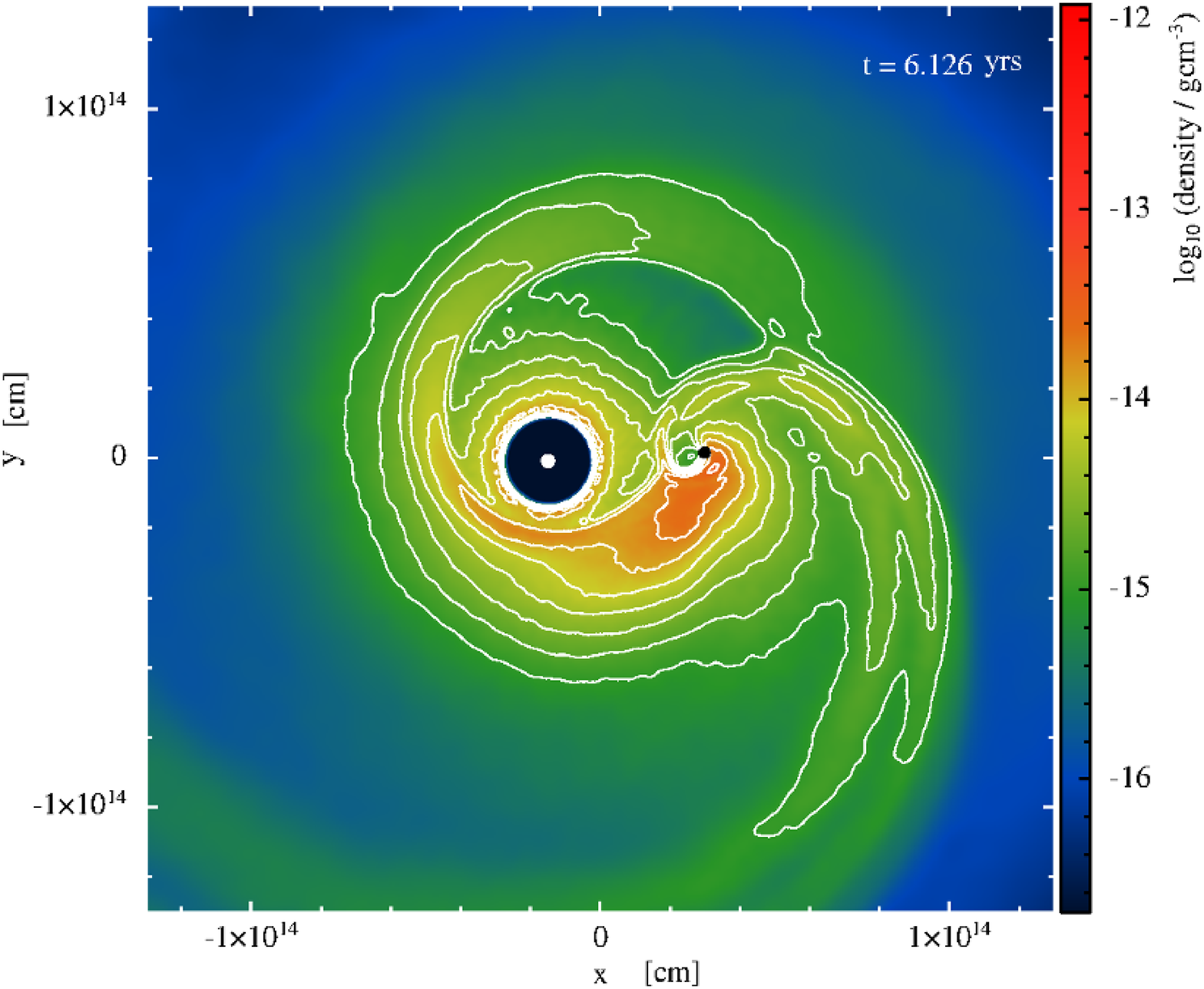}}
    {\includegraphics[width=0.495\textwidth, angle=360]{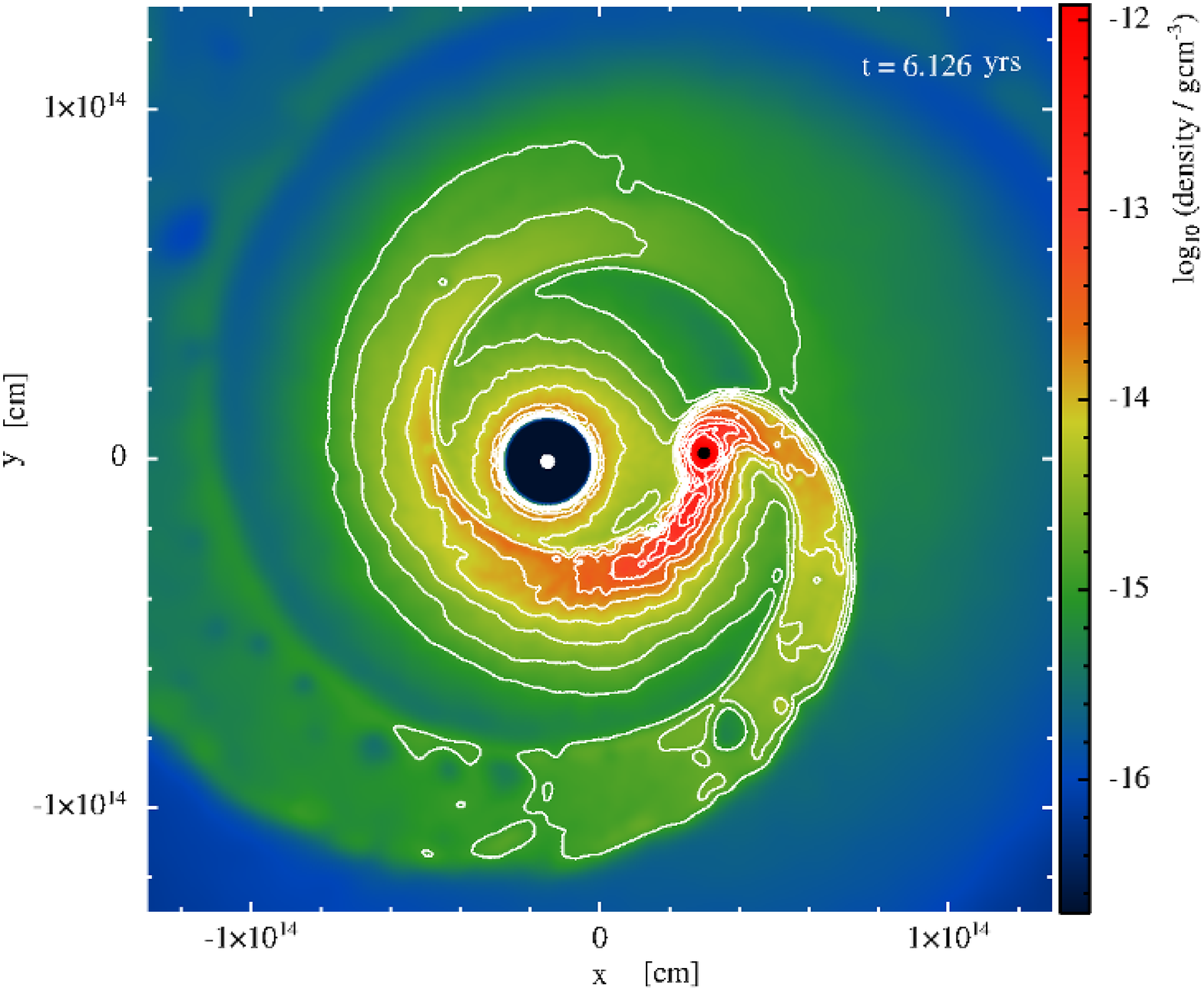}}
  \caption{Left panel: As Fig.~\ref{Fig:xy1}, but zoomed in on the binary system. Right panel: same as left panel, but for the isothermal $\gamma$=$1$ model. The contours show lines of the constant density. The white and black solid circle markers represent the mass-losing star and the accreting companion star, respectively.}
\label{Fig:zoom}
  \end{center}
\end{figure*}

\begin{figure*}
  \begin{center}
    {\includegraphics[width=0.495\textwidth, angle=360]{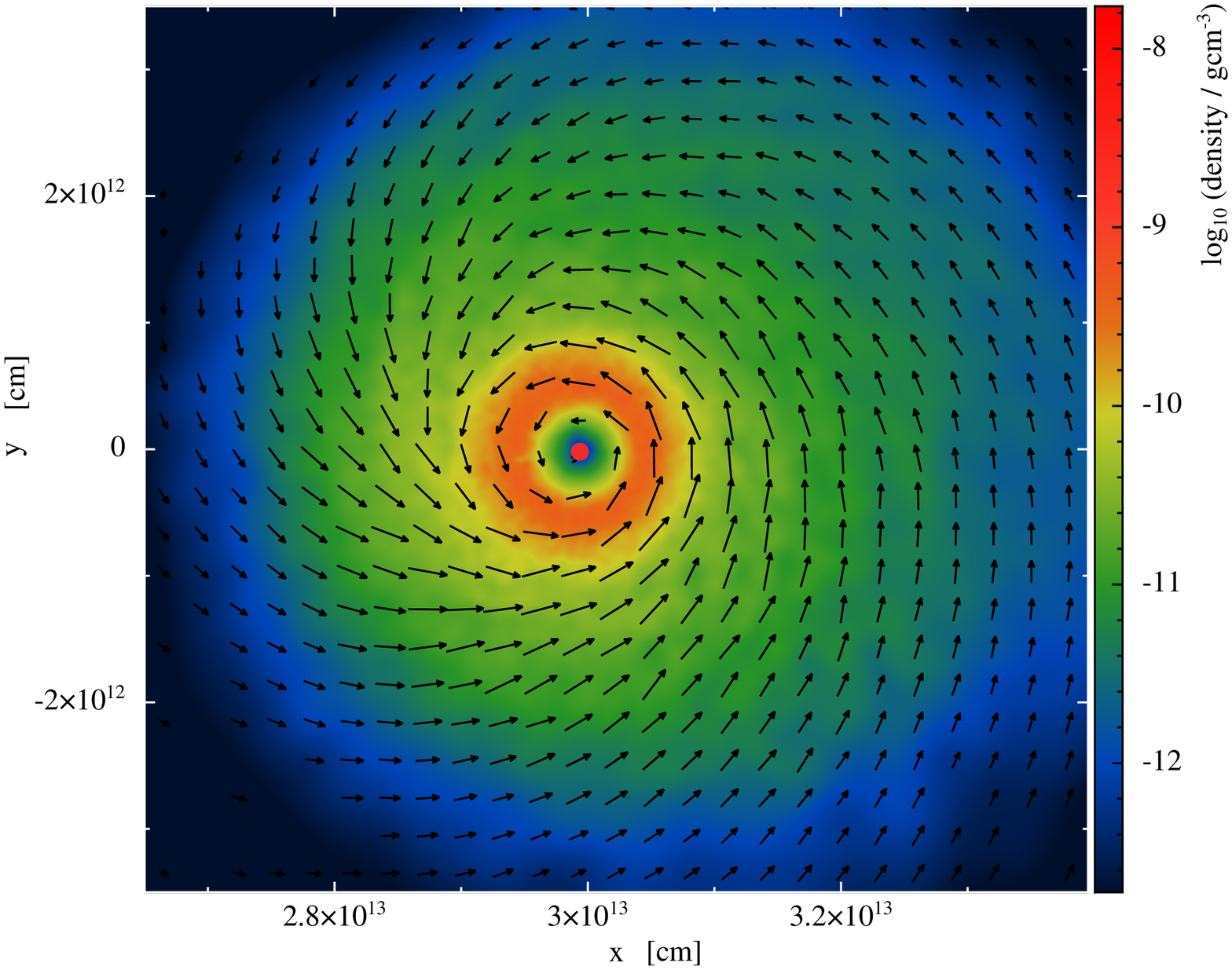}}
    {\includegraphics[width=0.495\textwidth, angle=360]{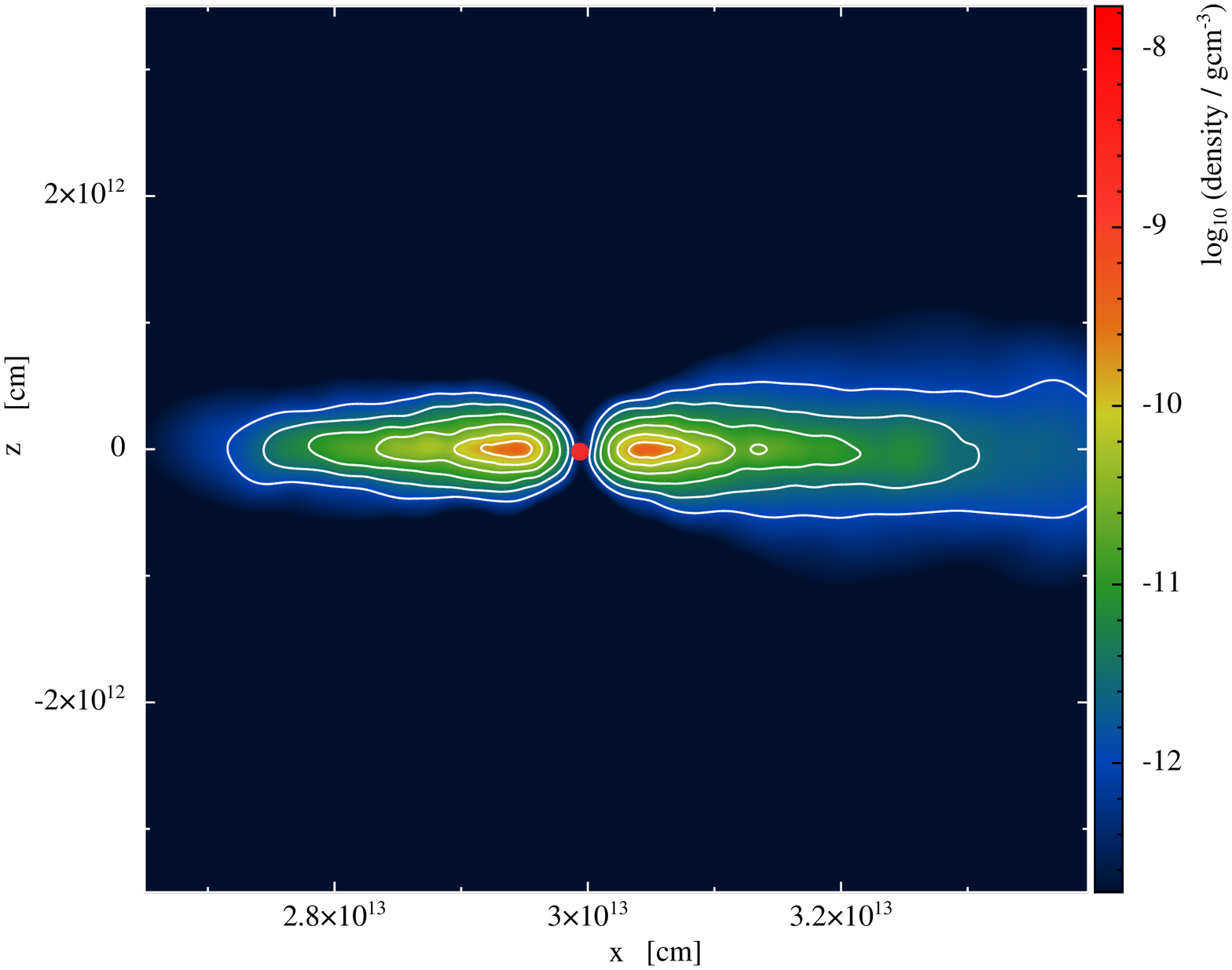}}
  \caption{As the right panel of Fig.~\ref{Fig:zoom}, but zoomed in on the accreting star (red solid circle) in the $x-y$ plane (left panel) and $x-z$ plane (right panel). In the left panel vectors are showing flow velocities. An accretion disc is clearly visible around the accreting star.}
\label{Fig:disc}
  \end{center}
\end{figure*}

Fig.~\ref{Fig:fit} shows the mass accretion efficiency and accreted specific angular momentum as a function of the mass ratio. We find that the mass accretion efficiency, $\eta_{\mathrm{acc}}$, generally increases with the mass ratio of a binary system, $q$=$M_2/M_1$. This is to be expected because, for fixed $M_{1}$ and $A_{\mathrm{orb}}$, as $M_{2}$ increases the gravitational attraction of the secondary becomes stronger and therefore more wind particles are attracted into its potential. However, the observed relation between $\eta_{\mathrm{acc}}$ and $q$=$M_2/M_1$ is more complicated. 
An empirical fit to the detailed results of the simulations is given by the following fifth order polynomial (left-hand panel of Fig.~\ref{Fig:fit}):
 \begin{equation}
\begin{split}
    \label{eq:3}
      \eta_{\mathrm{acc}} \approx -197.3\,q^{5} + 455.5\,q^{4}-355.4\,q^{3}+111.8\,q^{2}
                \\-7.410\,q+0.2649\,\hspace{64pt}\ \ \ \ \  [\,\%\,],
\end{split}
   \end{equation}

The dependence of the accreted specific angular momentum, $j_{\mathrm{acc}}$, on the mass ratio can be empirically fit by a fifth order polynomial (right-hand panel of Fig.~\ref{Fig:fit}):
 \begin{equation}
\begin{split}
    \label{eq:4}
      j_{\mathrm{acc}} \approx -2.657\,q^{5} + 6.642\,q^{4} - 5.810\,q^{3}+ 2.051\,q^{2}\hspace{0pt}
\\\,\,-0.1075\,q + 0.1185\hspace{17pt}\ \ \ \ \ [\,10^{19}\,\mathrm{cm^{2}\,s^{-1}}\,].
\end{split}
   \end{equation}

We point out that the above two non-linear relations are given to provide the best fits to the results of our simulations. They evade a simple physical interpretation. It might partly be explained by the differences of flow structures, which become more complex as the mass ratios of the systems increase (Fig.~\ref{Fig:xz}). Uncertainties owing to the simplified treatment of the wind acceleration mechanism (Section~\ref{sec:wind}) remain in our simulations and might play a role. Given these considerations, we therefore cannot exclude that the underlying relations are linear to a good approximation. Therefore, a linear fit to the results of our simulations is also given in Fig.~\ref{Fig:fit}. We point out that the above fitting equations are given based on the wind mass-transfer simulations without an accretion disc: The results are probably different if an accretion disc forms around the secondary (see discussion in Section~\ref{sec:gamma}).

\section{Discussions}
\label{sec:discussion}

\subsection{Convergence Test}
\label{sec:convergence}

To assess the reliability of our numerical results, we have performed a set of simulations for our default binary system with different resolutions by varying the mass of each SPH particle from $M_{\rm{SPH}}\approx10^{-10}\,M_{\sun}$ to $M_{\rm{SPH}}\approx3.0\times10^{-13}\,M_{\sun}$. Because all wind particles have the same mass, varying the mass of each particle corresponds to adjusting the total number of injected particles if the injection time is fixed. We found that the difference of the mass accretion efficiency ($\eta_{\mathrm{\,acc}}$) and the accreted specific angular momentum ($j_{\mathrm{\,acc}}$) between the resolution of $M_{\rm{SPH}}\approx3.6\times10^{-12}\,M_{\sun}$ and our highest resolution of $M_{\rm{SPH}}\approx3.0\times10^{-13}\,M_{\sun}$ is about 10\%. Again, considering that uncertainties on some physical processes such as the acceleration of the stellar wind due to the dust radiation remain in our simulations, we therefore think that our base resolution level of $M_{\rm{SPH}}\approx3.6\times10^{-12}\,M_{\sun}$ should be acceptable for the study of the present simulations.

\subsection{Simulations with a polytropic index of $\gamma$=$1$}
\label{sec:gamma}

To investigate how radiative cooling affects the results, simulations with various polytropic indices $\gamma$=$1, 1.1$ and 1.5 have been performed by past studies to represent more realistic situations that include the radiative cooling in the equation of state (e.g., \citealt{Theuns1996, Sato2003, de-Val-Borro2009}). It has been found that an accretion disc is clearly visible in the simulation for the case with $\gamma$=$1$ \citep{Theuns1996}. Here, we also run a simulation for an isothermal case with $\gamma$=$1$, which assumes that radiative cooling is such efficient that the gas temperature remains constant. Fig.~\ref{Fig:zoom} shows the comparison of flow structures between the case of $\gamma$=$5/3$ and the isothermal model with $\gamma$=$1$. The flow structure in the isothermal model is quite different from that in the model with $\gamma$=$5/3$. In particular, an accretion disc is clearly seen in the isothermal model (Fig.~\ref{Fig:disc}), which is dramatically different from the vortex structures seen in the case of $\gamma$=$5/3$ (Fig.~\ref{Fig:zoom}). As described in \citet{Theuns1993}, the reason for this difference might be the greater pressure of the gas in the case of $\gamma$=$5/3$ such that the gas stream is forced to expand vertically rather than confining to a disc as seen in the extreme case with  $\gamma$=$1$. In addition, like \citet{Theuns1993}, we also find the accretion disc to be fed from both the gas stream coming directly from the mass-losing star and the gas stream following the spiral arm behind the accreting star (see also \citealt{Theuns1996}). 

A model of the extreme case of $\gamma$=$1$ is set up for testing the role of the radiative cooling and heating of the gas in the simulation of wind mass transfer in binaries. This test shows that different choices of $\gamma$ have an influence on the results of the simulations. In our simulation with $\gamma$=$1$, we obtain that the mass accretion efficiency in this case is $\eta_{\mathrm{acc}}\approx11\%$, which is higher than the results (about 8\%) of \citet{Theuns1996}. However, as an accretion disc forms around the accreting star, the accretion of mass and angular momentum through the disc strongly depends on whether the disc is well resolved in the simulation, which is unlikely to be the case here. To test whether our discs are sufficiently well resolved requires more computing resources than are currently at our disposal. Therefore, in the present work, we particularly concentrate on the adiabatic simulations with a polytropic index $\gamma$=$5/3$. More realistic treatment of the radiative cooling and heating of the gas is needed to have a better physical understanding of the wind mass transfer in wide binaries. The effect of different $\gamma$ values on the structures of gas and the relationship between the mass ratio $q$ and mass accretion efficiency $\eta_{\mathrm{acc}}$ and angular momentum accretion rate $j_{\mathrm{acc}}$ are deferred to a forthcoming study.

\subsection{Implications of the Results}
\label{sec:impact}

A large fraction of wide binary stars, such as for example symbiotic systems, undergo mass and angular momentum exchange through wind mass transfer. This interaction has important consequences on the dynamical and chemical evolution of the binary systems, and therefore understanding the details of wind mass transfer is essential to determine their fates. For example, it is generally accepted that CEMP, barium-, and CH-stars are formed at different metallicities in binary systems in which the stellar wind of a thermally-pulsing asymptotic-giant branch (TPAGB) primary star is partly accreted by a lower-mass secondary star. The wind expelled by the primary star is enriched in the products of TPAGB nucleosynthesis, mainly carbon, barium and other $s$-process elements, and consequently the surface of the companion star is chemically enriched \cite[e.g.][]{McClure1984a, McClure1984b, McClure1990, Boffin1988, Han1995, Lucatello2005, Ryan2005}. The observable properties of these stars, in particular their surface abundances, are therefore strongly dependent on the transferred amount of mass and angular momentum. These two quantities are used as input parameters in detailed stellar-evolution studies investigating to what extent the accreted material is mixed throughout the secondary stars (e.g. \citealp{Stancliffe2007, Stacliffe2008, Matrozis2016, Matrozis2017}).

The accreted mass and angular momentum are key parameters also for population-synthesis studies which try to determine how many chemically-peculiar stars are formed for given initial distributions of primary and secondary masses, and orbital periods \cite[e.g.][]{Izzard2009, Izzard2010, Pols2012, Abate2013, Abate2015c}. The results of our simulations can be used as input to improve the parametric prescriptions currently adopted in population-synthesis models to determine the amount of mass and angular momentum that are accreted or lost by the binary systems \cite[cf. e.g.][]{Izzard2010, Abate2013, Abate2015a, Abate2015b, Abate2015c} and have a strong impact on the final results. For example, the accretion efficiency found in our default M07 model is $\eta_{\mathrm{acc}}\approx2\%$, that is about ten times smaller than the predictions of the canonical BHL model for a system with the same initial parameters. At present binary population-synthesis codes have to assume a more efficient wind-accretion model than the canonical BHL prescription to be able to reproduce the observed fraction of CEMP stars \cite[][]{Abate2015c}. Consequently, reducing the accretion efficiency would likely decrease the fraction of CEMP stars predicted by the models and hence increase the discrepancy with the observations. However, a small mass accretion efficiency would not be problematic for barium stars \citep{Boffin1994}.

On the other hand, as the star accretes mass and angular momentum, it should spin up. If the star spins up to its critical rate, further accretion might cease until the excess angular momentum is lost (\citealt{Packet1981}, but see \citealt{Popham1991}). At this point the fate of the accreting star must depend on what happens to the angular momentum of the accreting material \citep{Deschamps2013}. We note that the specific angular momentum found in our simulations is comparable to the Keplerian value at the surface of CEMP stars ($j_{\mathrm{K}}$=$\sqrt{GMR}\simeq 0.2\times10^{19}\,\mathrm{cm^{2}\,s^{-1}}$). Accreting more than a few hundredths of a solar mass of such material requires some angular momentum loss to occur during the accretion \citep{Packet1981}. If the star cannot lose the excess angular momentum, the amount of mass accreted could be much lower than otherwise expected, leading to some difficulties in explaining the chemical enrichment of some barium and most CEMP stars (e.g. \citealt{Matrozis2017}).

The results of our simulations may have strong implications also in the context of SNe Ia. The origin of SNe Ia is debated and symbiotic binary systems with a carbon-oxygen white dwarf (WD) accreting material from a mass-losing red giant (or AGB) star have been suggested as potential progenitors of SNe Ia \citep{Ruiz-Lapuente1997, Hachisu1999, Chiotellis2012}. In this scenario, at present population-synthesis studies show that the contribution of AGB stars as donors is negligible \cite[][]{Claeys2014} and the low accretion efficiencies determined in our simulations seem to confirm this conclusion. Moreover, given their compactness, substantial angular momentum loss during accretion must occur for accretion onto WDs to take place. A more realistic treatment of the wind acceleration mechanism in AGB stars and of the radiative cooling of the ejected gas may lead to higher wind accretion rates and consequently increase the contribution of AGB donor stars to the total SNe Ia rate \citep{Abate2017}, but the angular momentum loss likely constitutes a separate problem. In addition, the accreted angular momentum could result in that the WD spins with a short period which leads to an increase of the critical explosion mass. If the critical mass is higher than the actual mass of the WD, the SN explosion could only occur after the WD increases its spin period with a specific spin-down time-scale \citep{DiStefano2011, Justham2011}.

More and more observations suggest that a binary evolution in a majority of systems is very important in the formation and shaping of PNe \citep{Nordhaus2006, Jones2017a, Jones2017b}. The results of our simulations show that the structure of the outflow can be significantly affected by the mass ratio of the binary system (Section~\ref{sec:mass_ratio}). For instance, a sprial structure in the $x-y$ plane is clearly visible in the simulations with a low mass ratio. However, the spiral structure is less neat and it is visible only in the central regions as the mass ratio increases (Figs.~\ref{Fig:xz} and~\ref{Fig:xy3}). Our results imply that binary interaction may play an important role in the formation and shaping of PNe. Comparing our results with the observations of PNe is expected to be helpful for understanding the evolution of AGB stars and the shaping of PNe. In particular, using the observational data from Atacama Large Millimeter/submillimeter Array (ALMA), combined with hydrodynamic simulations, \citet{Maercker2012} have recently concluded that R Sculptoris is a binary system in which an AGB donor star underwent a thermal pulse associated with large mass loss and that the outflow has a spiral structure caused by the presence of a binary companion. 

\subsection{Uncertainties and Future Work}
\label{sec:uncertainties}

In this work, instead of considering the detailed wind acceleration mechanism due to radiation pressure on dust, we simply assume that the radiation pressure is balanced by the gravitational attraction of the mass-losing star in our model. This is a first step aimed to quantify the individual contributions of different physical parameters to the wind mass-transfer process in binary systems. Further simulations with a more realistic wind mechanism as done by \citet{Toupin2015b} are still needed. We are currently improving our wind model to consistently include the description of the gas chemistry and estimate the opacity of the gas, treating in detail the formation of dust and molecule (e.g., \citealt{Gail1984, Hoefner2003, Woitke2006}). This will allow us to calculate the wind acceleration due to radiation pressure and the radiative cooling self-consistently in the simulation (e.g., \citealt{Bowen1988, Woitke2006}). Also, stellar pulsations are not taken into account in present work because of uncertainties in the pulsation period, although it is suggested that stellar pulsation is a basic factor in the WRLOF accretion \citep{Mohamed2012}. By including stellar pulsations and the improved treatments of the radiative acceleration and cooling, we will investigate more details of wind mass transfer in binary systems such as the WRLOF case suggested by \citet{Mohamed2012} in a forthcoming study.

In this work, we primarily investigate how the mass ratio of a binary system affects the mass accretion efficiency and accreted specific angular momentum by wind mass transfer in a circular orbital with a separation of $A_{\mathrm{orb}}$=$3.0\,\rm{au}$. To provide a realistic prescription of the wind mass transfer in binary systems that can be implemented into D stellar evolution and population synthesis codes such as {\sc STARS} \citep{Eggleton1971, Pols1995, Stancliffe2009}, {\sc BINSTAR} \citep{Siess2011, Siess2013, Davis2013}, {\sc MESA} \citep{Paxton2011} and {\sc binary\_c} \citep{Izzard2004, Izzard2006, Izzard2009}, a more comprehensive study of the effects of the binary system properties (separation, eccentricity and wind characteristics, e.g., \citealt{deValBorro2017}) on the accreted mass and angular momentum is still needed. In particular, investigating the effect of eccentricity might help us to gain insight on the empirical evidence that many barium stars are found in eccentric systems \citep{BonacicMarinovic2008, Davis2013}. Also, the spin of the accreting star and how the accreted material affects the system geometry, both the separation and orbital eccentricity, are ignored in current simulations.

\section{Summary}
\label{sec:summary}

In this work, we have performed 3D hydrodynamical simulations of wind mass transfer in binary systems with the SPH code {\sc GADGET-3}. The overall aim of this work is to address the dependence of mass and angular momentum accretion on the mass ratio of the binary system. We particularly focus on the case of when the wind velocity is of the same order as the orbital velocity. The main results of this work can be summarized as follows.

\begin{itemize}
\item[1)]  The mass accretion efficiency $\eta_{\mathrm{acc}}$ and accreted specific angular momentum $j_{\mathrm{acc}}$ increase with the mass ratio, $q$=$M_{2}/M_{1}$. In the adiabatic simulations (polytropic index $\gamma$=$5/3$), we find that the mass accretion efficiency ($\eta_{\mathrm{acc}}$) varies from about $0.1\%$ to $8\%$ for the mass ratio between $q$=$0.05$ and $q$=$1.0$.

\item[2)] We have determined analytical relations (Eqs.~\ref{eq:3} and~\ref{eq:4}) to describe the dependence of $\eta_{\mathrm{acc}}$ and $j_{\mathrm{acc}}$ on $q$ in the case of an ideal gas with $\gamma$=$5/3$.

\item[3)] The geometry of the gas during the wind mass transfer process is found to be modified by the mass ratio, $q$, in our simulations (Fig.~\ref{Fig:xz}). The asymmetry and turbulence of the gas become more and more significant as $q$ increases.

\item[4)] With a simplified treatment of the wind acceleration mechanism, the results (e.g., mass accretion efficiency and gas flow structures) from our simulations are found to be consistent with previous studies (e.g., \citealt{Theuns1996}).

\item[5)]  For the isothermal, ideal gas case ($\gamma$=$1$), we find the flow pattern to be quite different from the case of $\gamma$=$5/3$. An accretion disc is clearly visible around the accreting star in the isothermal model.

\end{itemize} 
 
\

\section*{Acknowledgments}

We thank the referee for useful suggestions that improved the paper. Z.-W.L is grateful to Athena Stacy, Ekaterina V. Filippova and R\"udiger Pakmor for their help with modifying the code at the early stage of this project. We thank Jean-Claude Passy, Onno Pols and Martha Irene Saladino for many fruitful discussions. Some plots in the paper are made using the freely available SPLASH code \citep{Price2007}. C.A. acknowledges funding from the Alexander von Humboldt Foundation. R.J.S. is the recipient of a Sofja Kovalevskaja Award from the Alexander von Humboldt Foundation.

\bibliographystyle{aasjournal}

\bibliography{ref}

\begin{thebibliography}{}
\expandafter\ifx\csname natexlab\endcsname\relax\def\natexlab#1{#1}\fi

\bibitem[{{Abate}(2017)}]{Abate2017}
{Abate}, C. 2017, ArXiv e-prints, arXiv:1707.07988

\bibitem[{{Abate} {et~al.}(2015{\natexlab{a}}){Abate}, {Pols}, {Izzard}, \&
  {Karakas}}]{Abate2015b}
{Abate}, C., {Pols}, O.~R., {Izzard}, R.~G., \& {Karakas}, A.~I.
  2015{\natexlab{a}}, \aap, 581, A22

\bibitem[{{Abate} {et~al.}(2013){Abate}, {Pols}, {Izzard}, {Mohamed}, \& {de
  Mink}}]{Abate2013}
{Abate}, C., {Pols}, O.~R., {Izzard}, R.~G., {Mohamed}, S.~S., \& {de Mink},
  S.~E. 2013, \aap, 552, A26

\bibitem[{{Abate} {et~al.}(2015{\natexlab{b}}){Abate}, {Pols}, {Karakas}, \&
  {Izzard}}]{Abate2015a}
{Abate}, C., {Pols}, O.~R., {Karakas}, A.~I., \& {Izzard}, R.~G.
  2015{\natexlab{b}}, \aap, 576, A118

\bibitem[{{Abate} {et~al.}(2015{\natexlab{c}}){Abate}, {Pols}, {Stancliffe},
  {Izzard}, {Karakas}, {Beers}, \& {Lee}}]{Abate2015c}
{Abate}, C., {Pols}, O.~R., {Stancliffe}, R.~J., {et~al.} 2015{\natexlab{c}},
  \aap, 581, A62

\bibitem[{{Anzer} {et~al.}(1987){Anzer}, {Boerner}, \& {Monaghan}}]{Anzer1987}
{Anzer}, U., {Boerner}, G., \& {Monaghan}, J.~J. 1987, \aap, 176, 235

\bibitem[{{Bate} {et~al.}(1995){Bate}, {Bonnell}, \& {Price}}]{Bate1995}
{Bate}, M.~R., {Bonnell}, I.~A., \& {Price}, N.~M. 1995, \mnras, 277, 362

\bibitem[{{Benz}(1990)}]{Benz1990}
{Benz}, W. 1990, in Numerical Modelling of Nonlinear Stellar Pulsations
  Problems and Prospects, ed. J.~R. {Buchler}, 269

\bibitem[{{Boffin} \& {Jorissen}(1988)}]{Boffin1988}
{Boffin}, H.~M.~J., \& {Jorissen}, A. 1988, \aap, 205, 155

\bibitem[{{Boffin} \& {Zacs}(1994)}]{Boffin1994}
{Boffin}, H.~M.~J., \& {Zacs}, L. 1994, \aap, 291, 811

\bibitem[{{Bona{\v c}i{\'c} Marinovi{\'c}} {et~al.}(2008){Bona{\v c}i{\'c}
  Marinovi{\'c}}, {Glebbeek}, \& {Pols}}]{BonacicMarinovic2008}
{Bona{\v c}i{\'c} Marinovi{\'c}}, A.~A., {Glebbeek}, E., \& {Pols}, O.~R. 2008,
  \aap, 480, 797

\bibitem[{{Bondi}(1952)}]{Bondi1952}
{Bondi}, H. 1952, \mnras, 112, 195

\bibitem[{{Bondi} \& {Hoyle}(1944)}]{Bondi1944}
{Bondi}, H., \& {Hoyle}, F. 1944, \mnras, 104, 273

\bibitem[{{Bowen}(1988)}]{Bowen1988}
{Bowen}, G.~H. 1988, \apj, 329, 299

\bibitem[{{Bowen} \& {Willson}(1991)}]{Bowen1991}
{Bowen}, G.~H., \& {Willson}, L.~A. 1991, \apjl, 375, L53

\bibitem[{{Chen} {et~al.}(2017){Chen}, {Frank}, {Blackman}, {Nordhaus}, \&
  {Carroll-Nellenback}}]{Chen2017}
{Chen}, Z., {Frank}, A., {Blackman}, E.~G., {Nordhaus}, J., \&
  {Carroll-Nellenback}, J. 2017, ArXiv e-prints, arXiv:1702.06160

\bibitem[{{Chiotellis} {et~al.}(2012){Chiotellis}, {Schure}, \&
  {Vink}}]{Chiotellis2012}
{Chiotellis}, A., {Schure}, K.~M., \& {Vink}, J. 2012, \aap, 537, A139

\bibitem[{{Claeys} {et~al.}(2014){Claeys}, {Pols}, {Izzard}, {Vink}, \&
  {Verbunt}}]{Claeys2014}
{Claeys}, J.~S.~W., {Pols}, O.~R., {Izzard}, R.~G., {Vink}, J., \& {Verbunt},
  F.~W.~M. 2014, \aap, 563, A83

\bibitem[{{Davis} {et~al.}(2013){Davis}, {Siess}, \& {Deschamps}}]{Davis2013}
{Davis}, P.~J., {Siess}, L., \& {Deschamps}, R. 2013, \aap, 556, A4

\bibitem[{{De Marco} \& {Soker}(2011)}]{deMarco2011}
{De Marco}, O., \& {Soker}, N. 2011, \pasp, 123, 402

\bibitem[{{de Val-Borro} {et~al.}(2009){de Val-Borro}, {Karovska}, \&
  {Sasselov}}]{de-Val-Borro2009}
{de Val-Borro}, M., {Karovska}, M., \& {Sasselov}, D. 2009, \apj, 700, 1148

\bibitem[{{de Val-Borro} {et~al.}(2017){de Val-Borro}, {Karovska}, {Sasselov},
  \& {Stone}}]{deValBorro2017}
{de Val-Borro}, M., {Karovska}, M., {Sasselov}, D.~D., \& {Stone}, J.~M. 2017,
  ArXiv e-prints, arXiv:1704.03460

\bibitem[{{Deschamps} {et~al.}(2013){Deschamps}, {Siess}, {Davis}, \&
  {Jorissen}}]{Deschamps2013}
{Deschamps}, R., {Siess}, L., {Davis}, P.~J., \& {Jorissen}, A. 2013, \aap,
  557, A40

\bibitem[{{Di Stefano} {et~al.}(2011){Di Stefano}, {Voss}, \&
  {Claeys}}]{DiStefano2011}
{Di Stefano}, R., {Voss}, R., \& {Claeys}, J.~S.~W. 2011, \apjl, 738, L1

\bibitem[{{Duquennoy} \& {Mayor}(1991)}]{Duquennoy1991}
{Duquennoy}, A., \& {Mayor}, M. 1991, \aap, 248, 485

\bibitem[{{Edgar}(2004)}]{Edgar2004}
{Edgar}, R. 2004, \nar, 48, 843

\bibitem[{{Eggleton}(1971)}]{Eggleton1971}
{Eggleton}, P.~P. 1971, \mnras, 151, 351

\bibitem[{{Eggleton}(1983)}]{Eggleton1983}
---. 1983, \apj, 268, 368

\bibitem[{{Gail} {et~al.}(1984){Gail}, {Keller}, \& {Sedlmayr}}]{Gail1984}
{Gail}, H.-P., {Keller}, R., \& {Sedlmayr}, E. 1984, \aap, 133, 320

\bibitem[{{Gingold} \& {Monaghan}(1977)}]{Gingold1977}
{Gingold}, R.~A., \& {Monaghan}, J.~J. 1977, \mnras, 181, 375

\bibitem[{{Hachisu} {et~al.}(1999){Hachisu}, {Kato}, \& {Nomoto}}]{Hachisu1999}
{Hachisu}, I., {Kato}, M., \& {Nomoto}, K. 1999, \apj, 522, 487

\bibitem[{{Han} {et~al.}(1995){Han}, {Eggleton}, {Podsiadlowski}, \&
  {Tout}}]{Han1995}
{Han}, Z., {Eggleton}, P.~P., {Podsiadlowski}, P., \& {Tout}, C.~A. 1995,
  \mnras, 277, 1443

\bibitem[{{Hill} \& {Willson}(1979)}]{Hill1979}
{Hill}, S.~J., \& {Willson}, L.~A. 1979, \apj, 229, 1029

\bibitem[{{H{\"o}fner} {et~al.}(2003){H{\"o}fner}, {Gautschy-Loidl}, {Aringer},
  \& {J{\o}rgensen}}]{Hoefner2003}
{H{\"o}fner}, S., {Gautschy-Loidl}, R., {Aringer}, B., \& {J{\o}rgensen}, U.~G.
  2003, \aap, 399, 589

\bibitem[{{Hoyle} \& {Lyttleton}(1939)}]{Hoyle1939}
{Hoyle}, F., \& {Lyttleton}, R.~A. 1939, Proceedings of the Cambridge
  Philosophical Society, 35, 405

\bibitem[{{Hunt}(1971)}]{Hunt1971}
{Hunt}, R. 1971, \mnras, 154, 141

\bibitem[{{Ivezic} \& {Elitzur}(1997)}]{Ivezic1997}
{Ivezic}, Z., \& {Elitzur}, M. 1997, \mnras, 287, 799

\bibitem[{{Izzard} {et~al.}(2010){Izzard}, {Dermine}, \& {Church}}]{Izzard2010}
{Izzard}, R.~G., {Dermine}, T., \& {Church}, R.~P. 2010, \aap, 523, A10

\bibitem[{{Izzard} {et~al.}(2006){Izzard}, {Dray}, {Karakas}, {Lugaro}, \&
  {Tout}}]{Izzard2006}
{Izzard}, R.~G., {Dray}, L.~M., {Karakas}, A.~I., {Lugaro}, M., \& {Tout},
  C.~A. 2006, \aap, 460, 565

\bibitem[{{Izzard} {et~al.}(2009){Izzard}, {Glebbeek}, {Stancliffe}, \&
  {Pols}}]{Izzard2009}
{Izzard}, R.~G., {Glebbeek}, E., {Stancliffe}, R.~J., \& {Pols}, O.~R. 2009,
  \aap, 508, 1359

\bibitem[{{Izzard} {et~al.}(2004){Izzard}, {Tout}, {Karakas}, \&
  {Pols}}]{Izzard2004}
{Izzard}, R.~G., {Tout}, C.~A., {Karakas}, A.~I., \& {Pols}, O.~R. 2004,
  \mnras, 350, 407

\bibitem[{{Jahanara} {et~al.}(2005){Jahanara}, {Mitsumoto}, {Oka}, {Matsuda},
  {Hachisu}, \& {Boffin}}]{Jahanara2005}
{Jahanara}, B., {Mitsumoto}, M., {Oka}, K., {et~al.} 2005, \aap, 441, 589

\bibitem[{{Jones} \& {Boffin}(2017)}]{Jones2017b}
{Jones}, D., \& {Boffin}, H.~M.~J. 2017, ArXiv e-prints, arXiv:1705.00283

\bibitem[{{Jones} {et~al.}(2017){Jones}, {Van Winckel}, {Aller}, {Exter}, \&
  {De Marco}}]{Jones2017a}
{Jones}, D., {Van Winckel}, H., {Aller}, A., {Exter}, K., \& {De Marco}, O.
  2017, \aap, 600, L9

\bibitem[{{Justham}(2011)}]{Justham2011}
{Justham}, S. 2011, \apjl, 730, L34

\bibitem[{{Karovska}(1997)}]{Karovska1997}
{Karovska}, M. 1997, Journal of the American Association of Variable Star
  Observers (JAAVSO), 25, 75

\bibitem[{{Karovska} {et~al.}(2005){Karovska}, {Schlegel}, {Hack}, {Raymond},
  \& {Wood}}]{Karovska2005}
{Karovska}, M., {Schlegel}, E., {Hack}, W., {Raymond}, J.~C., \& {Wood}, B.~E.
  2005, \apjl, 623, L137

\bibitem[{{Knapp} {et~al.}(1998){Knapp}, {Young}, {Lee}, \&
  {Jorissen}}]{Knapp1998}
{Knapp}, G.~R., {Young}, K., {Lee}, E., \& {Jorissen}, A. 1998, \apjs, 117, 209

\bibitem[{{Kopal}(1959)}]{Kopal1959}
{Kopal}, Z. 1959, {Close binary systems}

\bibitem[{{Liu} {et~al.}(2012){Liu}, {Pakmor}, {R{\"o}pke}, {Edelmann}, {Wang},
  {Kromer}, {Hillebrandt}, \& {Han}}]{Liu12}
{Liu}, Z.~W., {Pakmor}, R., {R{\"o}pke}, F.~K., {et~al.} 2012, \aap, 548, A2

\bibitem[{{Liu} {et~al.}(2013){Liu}, {Pakmor}, {Seitenzahl}, {Hillebrandt},
  {Kromer}, {R{\"o}pke}, {Edelmann}, {Taubenberger}, {Maeda}, {Wang}, \&
  {Han}}]{Liu2013}
{Liu}, Z.-W., {Pakmor}, R., {Seitenzahl}, I.~R., {et~al.} 2013, \apj, 774, 37

\bibitem[{{Livio} {et~al.}(1986){Livio}, {Soker}, {de Kool}, \&
  {Savonije}}]{Livio1986}
{Livio}, M., {Soker}, N., {de Kool}, M., \& {Savonije}, G.~J. 1986, \mnras,
  222, 235

\bibitem[{{Lucatello} {et~al.}(2005){Lucatello}, {Tsangarides}, {Beers},
  {Carretta}, {Gratton}, \& {Ryan}}]{Lucatello2005}
{Lucatello}, S., {Tsangarides}, S., {Beers}, T.~C., {et~al.} 2005, \apj, 625,
  825

\bibitem[{{Lucy}(1977)}]{Lucy1977}
{Lucy}, L.~B. 1977, \aj, 82, 1013

\bibitem[{{Maercker} {et~al.}(2012){Maercker}, {Mohamed}, {Vlemmings},
  {Ramstedt}, {Groenewegen}, {Humphreys}, {Kerschbaum}, {Lindqvist},
  {Olofsson}, {Paladini}, {Wittkowski}, {de Gregorio-Monsalvo}, \&
  {Nyman}}]{Maercker2012}
{Maercker}, M., {Mohamed}, S., {Vlemmings}, W.~H.~T., {et~al.} 2012, \nat, 490,
  232

\bibitem[{{Mastrodemos} \& {Morris}(1998)}]{Mastrodemos1998}
{Mastrodemos}, N., \& {Morris}, M. 1998, \apj, 497, 303

\bibitem[{{Matrozis} {et~al.}(2017){Matrozis}, {Abate}, \&
  {Stancliffe}}]{Matrozis2017}
{Matrozis}, E., {Abate}, C., \& {Stancliffe}, R.~J. 2017, ArXiv e-prints,
  arXiv:1707.08224

\bibitem[{{Matrozis} \& {Stancliffe}(2016)}]{Matrozis2016}
{Matrozis}, E., \& {Stancliffe}, R.~J. 2016, \aap, 592, A29

\bibitem[{{Matsuda} {et~al.}(1992){Matsuda}, {Ishii}, {Sekino}, {Sawada},
  {Shima}, {Livio}, \& {Anzer}}]{Matsuda1992}
{Matsuda}, T., {Ishii}, T., {Sekino}, N., {et~al.} 1992, \mnras, 255, 183

\bibitem[{{McClure}(1984{\natexlab{a}})}]{McClure1984a}
{McClure}, R.~D. 1984{\natexlab{a}}, \pasp, 96, 117

\bibitem[{{McClure}(1984{\natexlab{b}})}]{McClure1984b}
---. 1984{\natexlab{b}}, \apjl, 280, L31

\bibitem[{{McClure} \& {Woodsworth}(1990)}]{McClure1990}
{McClure}, R.~D., \& {Woodsworth}, A.~W. 1990, \apj, 352, 709

\bibitem[{{Mohamed} \& {Podsiadlowski}(2007)}]{Mohamed2007}
{Mohamed}, S., \& {Podsiadlowski}, P. 2007, in Astronomical Society of the
  Pacific Conference Series, Vol. 372, 15th European Workshop on White Dwarfs,
  ed. R.~{Napiwotzki} \& M.~R. {Burleigh}, 397

\bibitem[{{Mohamed} \& {Podsiadlowski}(2010)}]{Mohamed2010}
{Mohamed}, S., \& {Podsiadlowski}, P. 2010, in American Institute of Physics
  Conference Series, Vol. 1314, American Institute of Physics Conference
  Series, ed. V.~{Kalogera} \& M.~{van der Sluys}, 51--52

\bibitem[{{Mohamed} \& {Podsiadlowski}(2011)}]{Mohamed2011}
{Mohamed}, S., \& {Podsiadlowski}, P. 2011, in Astronomical Society of the
  Pacific Conference Series, Vol. 445, Why Galaxies Care about AGB Stars II:
  Shining Examples and Common Inhabitants, ed. F.~{Kerschbaum}, T.~{Lebzelter},
  \& R.~F. {Wing}, 355

\bibitem[{{Mohamed} \& {Podsiadlowski}(2012)}]{Mohamed2012}
{Mohamed}, S., \& {Podsiadlowski}, P. 2012, Baltic Astronomy, 21, 88

\bibitem[{{Nagae} {et~al.}(2004){Nagae}, {Oka}, {Matsuda}, {Fujiwara},
  {Hachisu}, \& {Boffin}}]{Nagae2004}
{Nagae}, T., {Oka}, K., {Matsuda}, T., {et~al.} 2004, \aap, 419, 335

\bibitem[{{Nordhaus} \& {Blackman}(2006)}]{Nordhaus2006}
{Nordhaus}, J., \& {Blackman}, E.~G. 2006, \mnras, 370, 2004

\bibitem[{{Packet}(1981)}]{Packet1981}
{Packet}, W. 1981, \aap, 102, 17

\bibitem[{{Paczy{\'n}ski}(1971)}]{Paczynski1971}
{Paczy{\'n}ski}, B. 1971, \araa, 9, 183

\bibitem[{{Pakmor} {et~al.}(2012){Pakmor}, {Edelmann}, {R{\"o}pke}, \&
  {Hillebrandt}}]{Pakmor2012}
{Pakmor}, R., {Edelmann}, P., {R{\"o}pke}, F.~K., \& {Hillebrandt}, W. 2012,
  \mnras, 424, 2222

\bibitem[{{Paxton} {et~al.}(2011){Paxton}, {Bildsten}, {Dotter}, {Herwig},
  {Lesaffre}, \& {Timmes}}]{Paxton2011}
{Paxton}, B., {Bildsten}, L., {Dotter}, A., {et~al.} 2011, \apjs, 192, 3

\bibitem[{{Pols} {et~al.}(2012){Pols}, {Izzard}, {Stancliffe}, \&
  {Glebbeek}}]{Pols2012}
{Pols}, O.~R., {Izzard}, R.~G., {Stancliffe}, R.~J., \& {Glebbeek}, E. 2012,
  \aap, 547, A76

\bibitem[{{Pols} {et~al.}(1995){Pols}, {Tout}, {Eggleton}, \& {Han}}]{Pols1995}
{Pols}, O.~R., {Tout}, C.~A., {Eggleton}, P.~P., \& {Han}, Z. 1995, \mnras,
  274, 964

\bibitem[{{Popham} \& {Narayan}(1991)}]{Popham1991}
{Popham}, R., \& {Narayan}, R. 1991, \apj, 370, 604

\bibitem[{{Price}(2007)}]{Price2007}
{Price}, D.~J. 2007, \pasa, 24, 159

\bibitem[{{Pringle} \& {Wade}(1985)}]{Pringle1985}
{Pringle}, J.~E., \& {Wade}, R.~A. 1985, {Interacting binary stars}

\bibitem[{{Raghavan} {et~al.}(2010){Raghavan}, {McAlister}, {Henry}, {Latham},
  {Marcy}, {Mason}, {Gies}, {White}, \& {ten Brummelaar}}]{Raghavan2010}
{Raghavan}, D., {McAlister}, H.~A., {Henry}, T.~J., {et~al.} 2010, \apjs, 190,
  1

\bibitem[{{Rosswog}(2009)}]{Rosswog2009}
{Rosswog}, S. 2009, \nar, 53, 78

\bibitem[{{Ruffert}(1994)}]{Ruffert1994a}
{Ruffert}, M. 1994, \aaps, 106

\bibitem[{{Ruiz-Lapuente} \& {et al.}(1997)}]{Ruiz-Lapuente1997}
{Ruiz-Lapuente}, P., \& {et al.} 1997, in NATO Advanced Science Institutes
  (ASI) Series C, Vol. 486, NATO Advanced Science Institutes (ASI) Series C,
  ed. P.~{Ruiz-Lapuente}, R.~{Canal}, \& J.~{Isern}, 205

\bibitem[{{Ryan} {et~al.}(2005){Ryan}, {Aoki}, {Norris}, \& {Beers}}]{Ryan2005}
{Ryan}, S.~G., {Aoki}, W., {Norris}, J.~E., \& {Beers}, T.~C. 2005, \apj, 635,
  349

\bibitem[{{Sato} {et~al.}(2003){Sato}, {Sawada}, \& {Ohnishi}}]{Sato2003}
{Sato}, J., {Sawada}, K., \& {Ohnishi}, N. 2003, \mnras, 342, 593

\bibitem[{{Sawada} \& {Matsuda}(1992)}]{Sawada1992}
{Sawada}, K., \& {Matsuda}, T. 1992, \mnras, 255, 17P

\bibitem[{{Siess} {et~al.}(2011){Siess}, {Izzard}, {Davis}, \&
  {Deschamps}}]{Siess2011}
{Siess}, L., {Izzard}, R.~G., {Davis}, P.~J., \& {Deschamps}, R. 2011, in
  Astronomical Society of the Pacific Conference Series, Vol. 447, Evolution of
  Compact Binaries, ed. L.~{Schmidtobreick}, M.~R. {Schreiber}, \&
  C.~{Tappert}, 339

\bibitem[{{Siess} {et~al.}(2013){Siess}, {Izzard}, {Davis}, \&
  {Deschamps}}]{Siess2013}
{Siess}, L., {Izzard}, R.~G., {Davis}, P.~J., \& {Deschamps}, R. 2013, \aap,
  550, A100

\bibitem[{{Smith} {et~al.}(1996){Smith}, {Cunha}, {Jorissen}, \&
  {Boffin}}]{Smith1996}
{Smith}, V.~V., {Cunha}, K., {Jorissen}, A., \& {Boffin}, H.~M.~J. 1996, \aap,
  315, 179

\bibitem[{{Springel}(2005)}]{Springel2005}
{Springel}, V. 2005, \mnras, 364, 1105

\bibitem[{{Springel}(2010)}]{Springel2010}
---. 2010, \araa, 48, 391

\bibitem[{{Stancliffe} \& {Eldridge}(2009)}]{Stancliffe2009}
{Stancliffe}, R.~J., \& {Eldridge}, J.~J. 2009, \mnras, 396, 1699

\bibitem[{{Stancliffe} \& {Glebbeek}(2008)}]{Stacliffe2008}
{Stancliffe}, R.~J., \& {Glebbeek}, E. 2008, \mnras, 389, 1828

\bibitem[{{Stancliffe} {et~al.}(2007){Stancliffe}, {Glebbeek}, {Izzard}, \&
  {Pols}}]{Stancliffe2007}
{Stancliffe}, R.~J., {Glebbeek}, E., {Izzard}, R.~G., \& {Pols}, O.~R. 2007,
  \aap, 464, L57

\bibitem[{{Theuns} {et~al.}(1996){Theuns}, {Boffin}, \&
  {Jorissen}}]{Theuns1996}
{Theuns}, T., {Boffin}, H.~M.~J., \& {Jorissen}, A. 1996, \mnras, 280, 1264

\bibitem[{{Theuns} \& {Jorissen}(1993)}]{Theuns1993}
{Theuns}, T., \& {Jorissen}, A. 1993, \mnras, 265, 946

\bibitem[{{Toupin} {et~al.}(2015{\natexlab{a}}){Toupin}, {Braun}, {Siess},
  {Jorissen}, {Gail}, \& {Price}}]{Toupin2015}
{Toupin}, S., {Braun}, K., {Siess}, L., {et~al.} 2015{\natexlab{a}}, in EAS
  Publications Series, Vol.~71, EAS Publications Series, 173--174

\bibitem[{{Toupin} {et~al.}(2015{\natexlab{b}}){Toupin}, {Braun}, {Siess},
  {Jorissen}, \& {Price}}]{Toupin2015b}
{Toupin}, S., {Braun}, K., {Siess}, L., {Jorissen}, A., \& {Price}, D.
  2015{\natexlab{b}}, in Astronomical Society of the Pacific Conference Series,
  Vol. 497, Why Galaxies Care about AGB Stars III: A Closer Look in Space and
  Time, ed. F.~{Kerschbaum}, R.~F. {Wing}, \& J.~{Hron}, 225

\bibitem[{{Vassiliadis} \& {Wood}(1993)}]{Vassiliadis1993}
{Vassiliadis}, E., \& {Wood}, P.~R. 1993, \apj, 413, 641

\bibitem[{{Wallerstein} \& {Knapp}(1998)}]{Wallerstein1998}
{Wallerstein}, G., \& {Knapp}, G.~R. 1998, \araa, 36, 369

\bibitem[{{Woitke}(2006)}]{Woitke2006}
{Woitke}, P. 2006, \aap, 452, 537

\bibitem[{{Wood}(1979)}]{Wood1979}
{Wood}, P.~R. 1979, \apj, 227, 220

\end{thebibliography}

\end{document}